\documentclass{aa}  
\usepackage{graphicx}
\usepackage{txfonts}
\usepackage[inline]{enumitem}
\usepackage{natbib}
\begin{document} 

\title{The X-ray binaries in M83: will any of them form gravitational \\ wave sources for LIGO/VIRGO/KAGRA?}
\author{I.Kotko \inst{1} \thanks{ikotko@camk.edu.pl}  \and K.Belczynski \inst{1} \thanks{chrisbelczynski@gmail.com} }

\institute{Nicolaus Copernicus Astronomical Center, Polish Academy of Sciences, 
Bartycka 18, 00-716 Warsaw, Poland}
\date{}

\abstract{
There are 214 X-ray point-sources ($L_{\rm X}>10^{35} \mathrm{erg/s}$) identified as X-ray binaries (XRBs) in the 
nearby spiral galaxy M83. 
Since XRBs are powered by accretion onto a neutron star or a black hole from a companion/donor star these 
systems are promising progenitors of merging double compact objects (DCOs): black hole - black hole (BH-BH), 
black hole - neutron star (BH-NS), or neutron star - neutron star (NS-NS) systems. The connection (i.e. 
XRBs evolving into DCOs) may provide some hints to the yet unanswered question: what is the origin of the 
LIGO/Virgo/KAGRA mergers? 
Available observations do not allow to determine what will be the final fate of the XRBs observed in M83. 
Yet, we can use evolutionary model of isolated binaries to reproduce the population of XRBs in M83 by 
matching model XRB numbers/types/luminosities to observations. Knowing the detailed properties of M83 
model XRBs (donor/accretor masses, their evolutionary ages and orbits) we follow their evolution to the
death of donor stars to check whether any merging DCOs are formed.\\ 
Although all merging DCOs in our isolated binary evolution model go through the XRB phase (defined as reaching 
X-ray luminosity from RLOF/wind accretion onto NS/BH above $10^{35}$ erg/s), only very few XRBs evolve 
to form merging (in Hubble time) DCOs. For M83 with its solar-like metallicity stars and continiuous 
star-formation we find that only $\sim 1-2\%$ of model XRBs evolve into merging DCOs depending on the 
adopted evolutionary physics. This is caused by 
{\em (i)} merger of donor star with compact object during common envelope phase,  
{\em (ii)} binary disruption at the supernova explosion of donor star,  
{\em (iii)} formation of a DCO on a wide orbit (merger time longer than Hubble time). 
}

\titlerunning{XRBs in M83: will any of them form gravitational wave sources for LVK?}
\authorrunning{I.Kotko \& K.Belczynski}
\maketitle

\section{Introduction}
At the end of the third observing run the LIGO-Virgo-KAGRA (LVK) collaboration reported $\sim90$ detections of gravitational wave signals from coalescing double compact objects (DCOs) \citep{Abbott21}. Among them the vast majority ($\sim83$) are double black holes (BH-BH) mergers, $5$ are classified as black hole - neutron star (BH-NS) mergers and only $2$ are the confirmed mergers of two neutron stars (NS-NS). 
There are several formation scenarios that explain the origin of the merging DCOs. The leading ones are: the dynamical interactions in the dense stellar systems \cite[e.g.][]{Zwart00_denseStellar_evolution, Mapelli2016, Rodriguez2018}, the evolution of the isolated binaries \citep[e.g.][]{Lipunov1997,Belczynski10,Olejak21}, the evolution of the isolated multiple systems \citep[e.g.][]{Toonen16_Multiple,Gomez2021_Multiple,Stegmann2022_multiple}, the chemically homogeneous evolution of the isolated binaries \citep[e.g.][]{MandeldeMink2016}, the mergers of binaries in active galactic nuclei \citep{Antonini12,Fragione19,Tagawa20} and the primordial black holes \citep{deLuca20,CLESSE2022}. At present there is no conclusion what is the relative contribution of each of these channels to the formation of the observed gravitational waves (GW) sources . 
The 4th observational run of LVK, which is planned to start in 2023 and is scheduled to collect the data for roughly a year, is expected to rise significantly the number of detections by over $\sim 200$ events \citep{Petrov22}.  The characteristics of these events may help to constrain which of the DCOs evolutionary models are the most plausible ones.\\
While the merger is the final moment of the DCOs life, the X-ray binaries (XRBs) are considered as the DCOs possible progenitors. XRBs are the close binary systems where the mass is transferred from the donor star and accreted onto the compact object (NS or BH) giving rise to the X-ray emission. Certain aspects of the evolution of the binary systems which enter the XRB phase still remain unclear. To constrain uncertainties in the binary stars physics leading to the formation of XRBs many groups use evolutionary codes to synthesize XRBs populations and compare them with existing observations \citep[e.g.][]{B08,Toonen14_BPS,Eldridge2016_BPASS,Tassos22_POSYDON}. The most straightforward comparison is done via X-ray luminosity function (XLF, cumulative distribution of X-ray luminosity of all sources in a given sample) of a given XRBs population (in one or more galaxies). In most studies this kind of analysis is done for the combined XLF from a larger sample of galaxies \citep[e.g.][]{Fragos13,Tzanavaris13,L19,Misra22}.
In contrast, we focus on one galaxy with a large number of XRBs. \\
M83 (NGC~5236) is a spiral galaxy at the distance of $4.66\,\mathrm{Mpc}$ seen almost face-on, at the angle $i=24^{\circ}$, with the total star mass estimated to $\mathrm{M}_{\star}=2.138\times10^{10}\,\mathrm{M}_{\odot}$ and average star formation rate $\mathrm{SFR}\approx2.5\,\mathrm{M}_{\odot}\,\mathrm{yr}^{-1}$ \citep[hereafter: L19]{L19}.\\
Our choice of the galaxy is motivated by two reasons: 
\begin{enumerate*}[label=(\roman*)]
\item{it has one of the highest numbers of the detected X-ray point-sources (see L19)}
\item{the X-ray point-source population was cleared out of supernova remnants (SNRs) and background AGNs, leaving only the XRBs confirmed through optical observations \citep[hereafter: H21]{H21}}.
\end{enumerate*}\\

\section{Classification of X-ray binaries}\label{Sec:XRB_class}
The most general classification distinguishes $3$ groups of XRBs according to the mass of the donor star $\mathrm{M}_2$. For the sake of consistency between our models and observations we adopt the limits for $\mathrm{M}_2$ defined in H21:
\begin{enumerate}[nosep]
    \item{low-mass X-ray binaries (LMXBs) - where $\mathrm{M}_2\leq3\,\mathrm{M}_{\odot}$}
    \item{intermediate-mass X-ray binaries (IMXBs) - where $
    3\,\mathrm{M}_{\odot}<\mathrm{M}_2<8\,\mathrm{M}_{\odot}$}
    \item{high-mass X-ray binaries  (HMXBs) - where $\mathrm{M}_2\geq8\,\mathrm{M}_{\odot}$}
\end{enumerate}
Within the HMXBs group one may separate further $3$ types with regard to the donor star type \citep[e.g.][]{KRETSCHMAR2019}:
\begin{enumerate*}[label=(\roman*)]
\item{HMBXs with O/B supergiant donor star}
\item{Wolf-Rayet X-ray Binaries, where the donor is a Wolf-Rayet star, and }
\item{Be-XRBs, where the donor is a rapidly spinning B-star showing Balmer emission lines.}
\end{enumerate*}
The characteristic feature of the latter HMBXs type is their transient nature. They show strong enhancements of X-ray emission which may last for weeks to months after which there follows a long quiescent phase (of order of years). This behaviour is believed to be driven by the temporal accretion onto the compact object when it passes though the decretion disc surrounding a Be-star. In this work we do not distinguish between different types of HMBXs, however, we mention them for the sake of further discussion (Sec.~\ref{Sec:Discussion}).\\
The accretion in XRBs occurs through two mechanisms: stellar winds and Roche lobe overflow (RLOF). Wind-fed accretion dominates for the systems in which donor is a massive star with strong winds while RLOF takes place when the donor overfills its Roche lobe and the mass transfer onto the accretor begins. The latter may happen for the large range of donor radii (i.e. donor evolutionary stages) depending on the binary components mass ratio and orbital separation. For HMXBs the dominant accretion source is stellar wind while for LMXBs RLOF prevails. Although there are exceptions in both groups, e.g. Cen~X-3 which is HMXBs in which accretion through RLOF has been observed \citep{Sanjurjo_Ferr2021,Tsygankov22} or symbiotic X-ray binaries (SyXBs) which are considered as a subclass of LMXBs in which NS accrets matter from stellar wind of a late-type giant star \citep{Yungelson19,Lu12}.
To decide to which group a given XRB belongs one needs to complement the X-ray data with the observations of the donor star in the longer wavelength ranges (optical, UV, IR).

\section{Observations of XRBs in M83}\label{Sec:Obs}
The observational sample of XRBs to which we refer in this paper consists of $214$ systems among which $30$ are classified as LMXBs, $64$ as IMXBs and $120$ as HMXBs. The details of the observations and classification are presented in the following subsections.
\subsection{X-ray observations}\label{subs:X}
 The present generation of X-ray space telescopes with unprecedented resolution (\textit{Chandra}) and effective area (\textit{XMM-Newton}) allows to resolve the X-ray point-sources of luminosities $\mathrm{L_{X}}>10^{40}\,\mathrm{erg\,s^{-1}}$ at distances larger than  $100\,\mathrm{Mpc}$ \citep{DistXray}. At the distance of M83 the threshold luminosity for X-ray point-source detection is $L_X\approx10^{35}\,\mathrm{erg\,s^{-1}}$ at the exposure time of $\sim790\,\mathrm{ks}$ (\cite{Long14}). \\
The catalogue of M83 X-ray sources from \textit{Chandra} considered by L19 consists of $363$ objects laying within the isophotal ellipse that traces the $20\,\mathrm{mag\,arcsec^{-2}}$  galactic surface brightness in $K_s$-band. The ellipse semi-major and semi-minor axis are $a=5.21\,\mathrm{arcmin}$ and $b=4.01\,\mathrm{arcmin}$ respectively, giving the total area of $\sim 65.6\, \mathrm{arcmin}^2$. 
This catalogue has not been cleared from possible contamination from X-ray sources other than XRBs like SNRs or background X-ray sources like AGNs or quasars.

\subsection{Optical observations}\label{subs:optical}
 In the case of M83 the identification of XRBs was possible owing to the available optical observations of the galaxy taken by Hubble Space Telescope (HST) \citep{HST_M83}.\\
The area of M83 lying within the \textit{HST} footprint is $\sim\, 43 \mathrm{arcmin}^{2}$  \citep{H21} and contains $324$ X-ray point-sources from L19 catalogue.  Among them, using catalogues cross-referencing and analysis of the optical characteristics of the sources, H21 identified $103$ SNRs and $7$ background AGNs/quasars candidates. All remaining sources, i.e. $214$,  has been confirmed as XRBs. The regions not seen in the optical lay on the periphery of the galaxy. \\
The identification of XRBs based on the donors mass was done by H21 by comparing their position on the color-magnitude diagram with theoretical evolutionary mass tracks for solar metallicity stars from Padova models. From this method one can derive that there are $30$ LMXBs, $64$ IMXBs and $120$ HMXBs in the observed XRBs population. For details see H21. \\
For our analysis it is also relevant to consider the spatial distribution of XRBs in M83. From maps presented in H21 it can be seen that most of HMXBs are located in the spiral arms and in the galaxy buldge while LMXBs and IMXBs are more evenly distributed throughout the galaxy disc.  

\section{Population synthesis calculations}\label{Sec:Method}
To generate the intrinsic population of XRBs in M83 we use {\fontfamily{cmtt}\selectfont {\fontfamily{cmtt}\selectfont StarTrack}} population synthesis code  \citep{B02,B08,Belczynski20} which has been developed over the years. We adopt the 3-broken power law initial mass function (IMF) with exponents: $\alpha_1=-1.3$ for stars of masses $0.08<m<0.5\,\mathrm{M}_\odot$, $\alpha_2=-2.2$ for $0.5\leq m<1.0\,\mathrm{M}_\odot$ and $\alpha_3=-2.3$ for stars more massive than $1\,\mathrm{M}_\odot$ \citep[see][]{Kroupa93,Kroupa01}.  To generate the population of primordial binaries we draw the mass of the initially more massive star (the primary) $M_1$ from the mass distribution described by IMF within the range $M_1\in[5-150]\,\mathrm{M}_{\odot}$. The mass of the secondary (initially less massive star) $M_2$ is derived from the mass ratio $q=\frac{M_2}{M_1}$ for which we assume the flat distribution $\Phi(q)=1$ in the range $q=0-1$. The minimum for $M_2$ is set to $0.08\,\mathrm{M}_{\odot}$ which is the limiting mass for the onset of hydrogen burning in the star. The orbital period  ($P/\mathrm{d}$) (in units of days) and eccentricity $e$ are drawn from the distributions:  $f(\log{P/\mathrm{d}})=(\log{P/\mathrm{d}})^{-0.55}$  with $\log{P/\mathrm{d}}$ in the range $[0.15,5.5]$ and $f(e)=e^{-0.42}$ with $e$ in the range $[0.0,0.9]$ respectively. Both distributions are adopted from \citet{Sana12} and \citet{Sana13} with extrapolation of orbital periods to $\log{P/\mathrm{d}}=5.5$ following \citet{deMink15}.
\\
In our calculations we employ delayed supernova engine \citep{Fryer12} resulting in the continuous mass distribution of the compact objects, i.e. with no gap in remnant masses between $\sim2\,\mathrm{M}_{\odot}$ and $\sim5\,\mathrm{M}_{\odot}$. For the winds of low- and intermediate-mass stars the code follows the formulae of \citet{Hurley00} while the stellar winds of massive (O/B type) stars are described by the formulae for radiation driven mass loss from \citep{Vink2001} with inclusion of Luminous Blue Variable mass loss \citep{Belczynski10}. In the case of Wolf-Rayet stars winds we take into account the metallicity dependence \citep{Vink_deKoter05} and
the inhomogeneities (clumping) in the winds \citep{Hamann_Koesterke98}. \\
The natal kick velocity magnitude is drawn from the Maxwellian distribution with $\sigma=265\,\mathrm{km\,s^{-1}}$  \citep{Hobbs05} and is decreased by the fraction of the stellar envelope that falls back during the final collapse of the star, after being initially ejected in SN explosion. This fraction is defined by the fallback parameter $f_{\mathrm{fb}}$ which takes values from $0$ to $1$. The directions of the kicks are assumed to be distributed isotropically.
The code also accounts for the natal kicks of NS born through electron supernova (ECS) or accretion induced collapse (AIC) of a white dwarf (WD). We take their magnitude to be $20\%$ of the magnitude of core-collapse SN kicks.
In the case of direct (without SN explosion) BH formation from the most massive stars ($M_\mathrm{ZAMS}>42\,M_{\odot}$) we assume that BH receives no natal kick.\\
RLOF may be stable or unstable depending on the timescale at which the mass is being transferred. If the donor is transferring mass at the timescale longer than its thermal timescale it is able to maintain its thermal equilibrium and RLOF is stable on nuclear timescale. Otherwise, RLOF may lead to the thermal timescale mass transfer (TTMT) or to the dynamical instability resulting in the common envelope (CE) phase. In what we call standard model we apply the standard {\fontfamily{cmtt}\selectfont StarTrack} criteria for CE development described in \citet{B08} that lead to the formation of merging DCOs mostly through CE evolution (Model 1 and 2). \\
From recent stellar models and close binaries simulations \citep[e.g.][]{Pavlovskii2017_CE,Misra20,Ge20} it has been realized that the mass transfer from a donor star may be stable for much wider parameter space than it is with the standard assumptions for CE development. To take this results into account the alternative CE development criteria have been introduced in {\fontfamily{cmtt}\selectfont StarTrack} by \citet{Olejak21_CE}. According to these criteria CE develops if all below are fulfilled:
{\em (i)} the donor is an H-rich envelope giant during RLOF, 
{\em (ii)} the mass of the donor is $M_{\mathrm{ZAMS,don}}>18\,\mathrm{M}_{\odot}$,
{\em (iii)} the ratio of the accretor mass to the donor mass is smaller than the limiting value $q_\mathrm{CE}$ depending on the metallicity and donor mass (the values of $q_\mathrm{CE}$ are listed in Eq.$2$ and $3$ in \citet{Olejak21_CE}),
{\em (iv)} the donor radius at the onset of RLOF is in the regime where the expansion or convection instability appears 
\citep[see][for details]{Pavlovskii2017_CE,Olejak21_CE}.
It occurs that these criteria allow for the formation of the majority of merging BH-BHs through the stable RLOF channel only. \\
To see how the alternative CE development criteria impact the XRBs population that we consider and the number of mergining DCOs originating from it we calculate the model which differs from the standard Model 2 only in the alternative criteria for CE development applied (Model 3).\\

In our approach the stable RLOF onto compact object is Eddington limited. If the mass transfer rate exceeds the Eddington limit for the accretion rate only the fraction $f_\mathrm{a}$ of mass is accreted onto CO while the remaining material $(1-f_\mathrm{a}$ is expelled from the system (non-conservative RLOF) with the specific angular momentum of the accretor. When mass transfer rate is lower than the critical value the conservative RLOF is assumed ($f_\mathrm{a}=1$).
In the case of an accretion onto the non-degenerated star we adopt the non-conservative RLOF with $50\%$ of the mass transferred from the secondary being accreted and the remaining mass leaving the system with its associated angular momentum which is a combination of orbital and donor angular momenta \citep{B08}. \\
The common envelope stage is treated within the energy formalism \citep{Webbink84} which compares the envelope's energy with the change of the orbital energy. In this prescription there are two parameters introduced: $\lambda$ describing the donor central concentration and $\alpha_{\mathrm{ce}}$ parameterizing the efficiency at which the orbital energy is transferred into the envelope. In our calculations we use one CE parameter defined as $\alpha_\mathrm{CE}=\alpha_{\mathrm{ce}}\times\lambda$. We assume that $\alpha_{\mathrm{ce}}=1.0$ and for $\lambda$ we use the fits from \citet{Xu2010,Xu2010_Erratum}. We assume the pessimistic scenario where we do not allow CE survival for Hertzsprung gap donors as these stars are in the phase of rapid expansion and they have not developed the clear boundary between core and envelope yet and all such systems are expected to merge \citep{Belczynski07}.\\
For the stages in the binary evolution when accretion onto a compact object takes place through stable RLOF or stellar winds, we apply the accretion model described in \citet{Mondal}. The possible beaming effect in the case of the super-Eddington luminosities is accounted for following the prescription from \citet{King09}. 
\\
To account for Chandra observational band ($0.5-8\,\mathrm{keV}$) we apply the bolometric correction (BC) to the bolometric luminosity calculated in the code. We use the detailed calculations from \citet{A22} where the bolometric correction depends on the mass accretion rate and the type of an accretor. \\
The observations show that there is a gradient of metallicity in M83: it has the value as high as $Z\sim0.03$ in the centre and it falls down to $Z\sim0.006$ in the far outer disc \citep{Bresolin05,Bresolin09}. We adopted the value $Z=0.01$ as an average in the region covered by \textit{HST}, but we also run models for metallicities: $Z=0.02,0.025,0.03$.\\ The star formation rate of M83 derived by \citet{L19} is $\mathrm{SFR}=2.5\,\mathrm{M}_{\odot}\,\mathrm{yr}^{-1}$ and has the uncertainty of $0.1$ dex (i.e. $\sim1.26\,\mathrm{M}_{\odot}\,\mathrm{yr}^{-1}$). We checked $4$ values of SFR ($1.5,\,2.0,\,2.5,\,3.5\, \mathrm{M}_{\odot}\,\mathrm{yr}^{-1}$) laying within the uncertainty range. We also tested the effect of changing the power in IMF for massive stars from $-2.3$ to $-1.9$, i.e. within its uncertainty.

\section{Predicted number of XRBs in M83}\label{Sec:predicted_number_factors}
In order to constrain the calculations of a single model to the reasonable computational time (a few days) we limit the number of the systems that undergo evolution to $2\times10^7$. As a result the total stellar mass of a model is $\approx2.28\times10^9\,\mathrm{M}_{\odot}$ which is $\sim2$ orders of magnitude smaller than the mass of the galaxy of interest. The motivation to adopt continuous SFR comes from the observations of metal abundances and the long dynamical and crossing times in the extended ultraviolet (XUV) disk of M83 \citep{Bruzzese20}.  With known galaxy stellar mass $M_\star$ and known continuous star formation rate $SFR$  one can estimate age of the galaxy $t_{\mathrm{gal}}=M_\star/SFR$. To obtain the stellar mass of the synthetic population  of the stars $M_{\star,\mathrm{synt}}$ equal to the mass of M83 ($M_\star$) each system from the generated population of the binaries has to be used $9$ times ($M_{\star,\mathrm{synt}}\times9=M_\star$). Therefore, for each binary we draw its birth time $t_0$ nine times from the range $[0,t_{\mathrm{gal}}]$ and we add to $t_0$ the time it took for the system to evolve to the XRB stage $t_\mathrm{X,start}$. To select the present-day XRBs in the galaxy we choose only those binaries which cross the $t_{\mathrm{gal}}$ during their XRBs phase, i.e. the time interval from the onset of their XRB phase $t_0+t_\mathrm{X,start}$ to the end of their XRB phase $t_0+t_\mathrm{X,end}$ contains $t_{\mathrm{gal}}$. The present-day fraction of XRBs is the intrinsic number of XRBs ($N_{\mathrm{i}}$) in our model of galaxy M83.\\
 To obtained the population of $N_{\mathrm{i}}$ XRBs one needs several additional post-processing steps so that it can be compared with the XRBs observed in M83. Each of the steps described below leads to the consecutive reductions in the number of XRBs in our sample and for each step we give a specific factor $f_{\mathrm{x}}$ by which the intrinsic number of the synthetic XRBs has to be reduced. One can encapsulate this process in $3$ equations depending on the type of XRB: 

\begin{equation} \label{eq:Nxrb,LMXB}
N_{\mathrm{f,LMXB}}=f_{\mathrm{L_X}}\,f_{\mathrm{vis}}\,(f_{\mathrm{P}}+f_{\mathrm{DC}}-f_{\mathrm{P}}\,f_{\mathrm{DC}})\,N_{\mathrm{i,LMXB}}
\end{equation}
\begin{equation} \label{eq:Nxrb,IMXB}
N_{\mathrm{f,IMXB}}=f_{\mathrm{L_X}}\,f_{\mathrm{vis}}\,N_{\mathrm{i,IMXB}}
\end{equation}
\begin{equation} \label{eq:Nxrb,HMXB}
N_{\mathrm{f,HMXB}}=f_{\mathrm{L_X}}\,N_{\mathrm{i,HMXB}} ,
\end{equation}
where $N_{\mathrm{f,j}}$ is the predicted observed number of XRBs of type j=LMXB, IMXB or HMXB in M83 for a given model, $N_{\mathrm{i,j}}$ is the  intrinsic number of XRBs of a given type from the simulations and $f_{\mathrm{x}}$ are the reduction factors described below. The final number of all XRBs after the reduction in a given model is $N_{\mathrm{f,XRBs}}=N_{\mathrm{f,LMXB}}+N_{\mathrm{f,IMXB}}+N_{\mathrm{f,HMXB}}$. \\
The values of $f_{\mathrm{x}}$ given in the following subsections are calculated for Model $2$. The reduction factors for all models listed in Table~\ref{table:BestFit} are given in Table~\ref{Table:Models_factors}.

\subsection{The X-ray luminosity threshold - $f_\mathrm{L_X}$}

In accordance with the observational threshold we choose only the sources with the X-ray luminosity  $\mathrm{L_{X}}\geq 10^{35}\,\mathrm{erg\,s^{-1}}$ which causes that we are left with less than $27\%$ of the initial number of sources. The corresponding factor $\mathrm{f_\mathrm{x}}$ is $f_{\mathrm{L_X}}=0.271$.

\subsection{The spatial coverage of \textit{HST} observations of M83 - $f_\mathrm{vis}$} 

The XRBs population from {\fontfamily{cmtt}\selectfont StarTrack} corresponds to the XRBs population of the entire M83. To account for the difference between X-ray and optical observational spatial coverage of M83 we proceed with the following assumptions: 
\begin{enumerate*}[label=(\roman*)]
\item{the whole XRBs population of M83 is located within the area of X-ray observations described in Sec.~\ref{subs:X};}\label{item:fvis-ST}
\item{due to the spatial distribution of different types of XRBs (see Sec.~\ref{subs:optical}) all HMXBs are seen by \textit{HST} but some LMXBs and IMBXs may be located out of its footprint;}\label{item:fvis-distrib}
\item{$85\%$ of SNR originates from the core collapse SN (CC-SNR), i.e. has massive stars progenitors while $15\%$ comes from Ia SN (SNIa-SNR) and has low-mass stars progenitors \citep{Tammann94}, therefore, they follow the same spatial distribution as HMXBs and LMXBs accordingly.}\label{item:fvis_SNR}
\end{enumerate*}
From the assumption (iii) we find that there are $88$ CC-SNR and $15$ SNIa-SNR among $103$ SNR identified in optical (see Sec.~\ref{subs:optical}). Because SNIa-SNR are supposed to be approximately evenly distributed in the galaxy they number should scale proportionally to the observed area while the number of observed CC-SNR is the same both for X-ray and optical observations as they follow the high star formation rate regions of the galaxy. 
Taking this into account there should be $118$ non-XRBs among \textit{Chandra} X-ray sources: $88$ CC-SNRs, $23$ SNIa-SNRs and $7$ AGNs.
This means that $245$ out of $363$ X-ray sources could be identified as XRBs if \textit{HST} would cover the same area of M83 as \textit{Chandra} with $120$ being HMXBs and $125$ belonging to LMXB or IMXB type. Consequently there are $31$ LMXBs/IMXBs in M83 that are "lost" in the optical band due to \textit{HST} footprint and 
our reduction factor for the observational spatial coverage is $f_\mathrm{vis}=0.752$ with the caveat that we reduce only the number of LMXBs and IMXBs.

\subsection{The transient and persistent LMXBs - $f_{\mathrm{P}}$ and $f_{\mathrm{DC}}$}\label{subs:transients}

Many of LMXBs are transient in nature (T-LMXBs). They cycle between the short periods of high luminosity (outbursts) with peak luminosity of order $L_\mathrm{X}\sim10^{36}-10^{39}\,\mathrm{erg\,s^{-1}}$ and long periods of quiescence, when their luminosity is below $10^{33}\,\mathrm{erg\,s^{-1}}$. Those transitions are
explained by the thermal-viscous instability developing in the accretion disc \citep[see][]{Osaki74,Smak82,Lasota01,Hameury20}. According to the disc instability model (DIM) there is a critical value of the mass accretion rate ($\dot{M}_{\mathrm{acc,crit}}$) at the inner disc radius above which the system will stay permanently in the luminous state and one will observe it as a persistent LMXB (P-LMXBs). In the case of such close binaries as LMXBs (which orbital periods are in the range of tens of minutes to tens of days) the irradiation of the disc by the X-rays generated at its inner edge cannot be neglected. It gives the additional source of the heating of the disc and causes  $\dot{M}_{\mathrm{acc,crit}}$ to be lower.
For the critical values of mass accretion rate we use the formulae from \citet{Lasota08} for helium-rich irradiated disc and H-rich irradiated disc respectively:
\begin{equation}\label{eq:mcrit_He}
\dot{M}_{\mathrm{acc,crit,He}} =4.69\times10^{-12}\,C_{\mathrm{irr}}^{-0.22}\,f^{2.51}\,P_{\mathrm{orb}}^{1.67}\,M_2^{0.1}\,\,\mathrm{M_{\odot}\,yr^{-1}}
\end{equation}
\begin{equation}\label{eq:mcrit_solar}
 \dot{M}_{\mathrm{acc,crit,H}} =1.92\times10^{-13}\,C_{\mathrm{irr}}^{-0.36}\,f^{2.39}\,P_{\mathrm{orb}}^{1.59}\,M_2^{0.16}\,\,\mathrm{M_{\odot}\,yr^{-1}}
 \end{equation}
 where $C_{\mathrm{irr}}$ is the disc-irradiation parameter in units of $0.001$, $f=0.6(1+q)^{-2/3}$, $q=M_2/M_1$ (see Sec.~\ref{Sec:Method}),  $P_{\mathrm{orb}}$ is an orbital period in minutes and $M_2$ is mass of a donor.\\
The most uncertain parameter in formulae \ref{eq:mcrit_He} and \ref{eq:mcrit_solar} is $C_{\mathrm{irr}}$ which describes the fraction of X-ray flux which heats up the disc. The attempts to constraint its value by observations of the outbursts in LMXBs has not been conclusive \citep[see][]{Tetarenko18}. Although the standard value used in the literature is $5.0\times10^{-3}$ we use $C_{\mathrm{irr}}=1.0\times10^{-3}$ which still gives the DIM predictions for the system stability consistent with the observed transient and persistent LMXBs \citep[e.g.][]{Coriat12}.
We classify LMXB in the model as persistent if the mass transfer rate from the donor is higher than $\dot{M}_{\mathrm{acc,crit}}$ calculated for the parameters of a given system and transient otherwise. We consider LMXBs in the model as observable if they are either P-LMXBs or T-LMXBs in outburst. 
The factor that reduces the number of LMXBs to the persistent systems only is calculated as a ratio of P-LMXBs to all LMXBs and is derived to be $f_{\mathrm{P}}=0.043$.\\
The duty cycle (DC) of a transient system is the fraction of its lifetime that the system spends in the outburst. \citet{Yan15} derived the value of duty cycle for all LMXBs equal to $0.025^{+0.045}_{-0.016}$ from the sample of $36$ systems observed by the \textit{Rossi X-ray Timing Explorer}. From there we adopt the value for the reduction factor $f_{\mathrm{DC}}$ which reduces the number of T-LMXBs to only those systems which are actually in outburst to be $f_{\mathrm{DC}}=0.025$.\\
Finally, we combine the above described $2$ reduction factors in Eq.\ref{eq:Nxrb,LMXB} in order to include all P-LMXBs and outbursting T-LMXBs in our predicted number of XRBs in M83.

\section{Results}\label{Sec:results}
\subsection{The model}\label{subs:models}
Out of all models with standard CE development criteria we tested, we chose $2$ based on the $3$ criteria: the XLF shape, the total number of XRBs and the numbers of XRBs in the particular subgroups: LMXBs, IMXBs and HMXBs. Model~1 is a model of XRB population that we get after applying all the reduction steps described in Sec.~\ref{Sec:predicted_number_factors} and it is a reference model to Model~2 which is the most satisfying model considering the aforementioned criteria. 
Additionally we calculated the model with the parameters of Model~2 ($\mathrm{SFR}=3.5\,\mathrm{M}_{\odot}\,\mathrm{yr}^{-1}$ and $Z=0.01$, standard IMF) but with revised CE development criteria. We refer to this model as Model~3.
 The model parameters and corresponding number of the systems for all three models can be found in Table~\ref{Table:Models_param} and Table~\ref{table:BestFit} respectively.\\
 We note that we change the limit for the maximum donor mass in LMXBs and minimum donor mass in IMXBs in Model 2 and 3. The motivation for thos change is discussed in  Sec.~\ref{Subsec:Dicuss_XRBs_number}.

 \begin{table*}
\caption{The parameters adopted for three models presented in the paper. The columns are: metallicity $\mathrm{Z}$, solar metallicity $\mathrm{Z}_{\odot}$, star formation rate, the index $\alpha_3$ of the initial mass function in the high mass stars regime, the adopted CE development criteria and the upper limit for a donor mass in LMXBs. In the last column we marked in which model we draw 30 LMXBs out of all synthetic LMXBs in a model (see Sec.~\ref{Subsec:Dicuss_XRBs_number})}.   % title of Table
\label{Table:Models_param}      % is used to refer this table in the text
                       % used for centering table
% \resizebox{\columnwidth}{!}{ 
\centering
\begin{tabular}{c |c c c c c c c}        % centered columns (4 columns)
\hline\hline                 % inserts double horizontal lines
\noalign{\smallskip}
Model & $\mathrm{Z}$ & $\mathrm{Z}_{\odot}$ & SFR & $\alpha_3$ (IMF) & CE dev. & $M_2$ limit & LMXBs\\
 & & & {\footnotesize{($\mathrm{M}_{\odot}\,\mathrm{yr}^{-1})$}} & {\footnotesize{($\mathrm{M}>1\,\mathrm{M}_{\odot})$}} & {\footnotesize{criteria}} & {\footnotesize in LMXBs} ($\mathrm{M}_{\odot}$) & draw \\ 
\noalign{\smallskip}
\hline                     % inserts single horizontal line
\noalign{\smallskip}
1 & 0.01 & 0.014 & 2.5 & -2.3 & standard & 3.0  & no\\      
2 & 0.01 & 0.014 & 3.5 & -2.3 & standard & 3.5 & yes\\
3 & 0.01 & 0.014 & 3.5 & -2.3 & alternative  & 3.5 & yes\\
   \hline  
\end{tabular}
\end{table*}

\subsection{XLF}\label{subs:results-XLF}
\subsubsection{The shape}\label{subs:XLF_shape}
One can use the cumulative XLF to check if the systems in the observational and synthetic populations have similar distributions. From the top to the bottom of Fig.~\ref{xlf} we plot the observational XLF (red stars) and the synthetic XLF (black solid line) for Model~1-3 accordingly. The shape of Model 1 and Model 2 XLFs is alike but the excess of the synthetic LMXBs in Model~1 causes that its XLF is shifted towards higher numbers on the vertical axis and larger luminosities on the horizontal axis. The higher number of HMXBs in the XRBs population and more luminous IMXBs in Model 3 than in Model 1 and 2 cause that its XLF slope is shallower up to the luminosity $\sim4.0\times10^{38}\,\mathrm{erg\,s^{-1}}$ and steeper above that luminosity than in the standard models.
\subsubsection{The number of binaries}\label{subsec:number_xrbs_group}
The number of the present-day XRBs is $N_{\mathrm{i}}=50\,834$ in Model~1, $N_{\mathrm{i}}=71\,164$ in Model~2 and $N_{\mathrm{i}}=75\,854$ in Model~3. In each of three models  about $\sim27\%$ of the binaries have the luminosities above $10^{35}\,\mathrm{erg\,s^{-1}}$. 
Following the reduction steps described in Sec.~\ref{Sec:predicted_number_factors} we get the synthetic 
population of the observable XRBs, $N_{\mathrm{f,XRBs}}$, consisting of 819 binaries in Model~1, $1130$ binaries in Model~2 and $1269$ binaries in Model~3.\\
We decompose XRBs populations into HMXBs, IMXBs, outbursting T-LMXBs and P-LMXBs and analyse how each of those contributes to the total XLF. By example of Model~1 it appears that the main culprit of the mismatch between the observations and the simulations are P-~LMXBs (cyan dashed line on the top panel of Fig.~\ref{xlf}) which are too numerous ($441$) and brighter than observed XRBs. There is also small overabundance of outbursting T-LMXBs (green dashed line on the top panel of  Fig.~\ref{xlf}) at luminosities $\log L_{\mathrm{X}}\in[35-35.5]\,\mathrm{erg\,s^{-1}}$. The number of outbursting T-LMXBs in Model~1 is $248$. We choose to make one more reduction step (additional to the reduction factors from Sec.~\ref{Sec:predicted_number_factors}) in Model~2 and 3, i.e. we draw $30$ LMXBs from the set consisting of P-LMXBs and outbursting LMXBs. As a result the final number of XRBs in Model~2 and 3 is $197$ and $263$ respectively compared to $214$ observed. The resulting XLFs are presented on the middle and bottom panel of Fig.~\ref{xlf}.

\subsubsection{The number of binaries in XRBs subgroups}\label{subsec:number_xrbs_subgroups}
What distinguishes  Model 2 (presented on the middle panel of Fig.~\ref{xlf}) from Models 1 and 3 and makes it our model of choice is not only the total number of XRBs but also the numbers of different types of XRBs which closely correspond to what is observed. From Model~2 we get $30$ LMXBs, $51$ IMXBs and $116$ HMXBs comparing to $30$ LMXBs, $64$ IMXBs and $120$ HMXBs observed. Although there are $30$ LMXBs and $64$ IMXBs in Model~3, what matches perfectly the observations, the number of synthetic HMXBs ($169$) exceeds the observed number by $\sim41\%$. \\

\subsubsection{The properties of synthetic XRBs}
To have more insight into out results we show the basic characteristic of XRBs populations in Model 1, 2 and 3. The  distributions of the donor masses are right-skewed in all 3 models (see magenta, light-green and light-blue boxes on left panels of Fig.~\ref{Fig:XRBs_mass_distrb}). The upper quartile of $M_\mathrm{donor}$ distribution in Model 1 is $\sim1.84\,M_{\odot}$ while is $\sim15.96\,M_{\odot}$ and $\sim16.54\,M_{\odot}$ in Model 2 and 3 respectively. This clearly demonstrates the drastic reduction of LMXBs to 30 systems in Model 2 and 3. \\ The distributions of the compact object (NS/BH) masses has a peak at $M_\mathrm{NS/BH}=1-2\,M_{\odot}$ corresponding to NSs type of the accreting object. The NSs constitute $\sim 77\%$ of the accretors in Model 1, in Model 2 the proportion between NSs and BHs is more balanced ($\sim 41\%$ NSs and $\sim 59\%$ BHs) while in Model 3 the NS/BH ratio is reverted comparing to Model 1 with $\sim 27\%$ NSs and $\sim 73\%$ BHs. The delayed SN model that we use in our evolution calculations (see Sec.~\ref{Sec:Method}) results in compact objects present is the so called "mass gap" ($\sim2-5\,M_{\odot}$) in $M_\mathrm{NS/BH}$ distribution of all 3 models.

\begin{figure}[ht!]
   \centering
   {\includegraphics[width=\columnwidth]{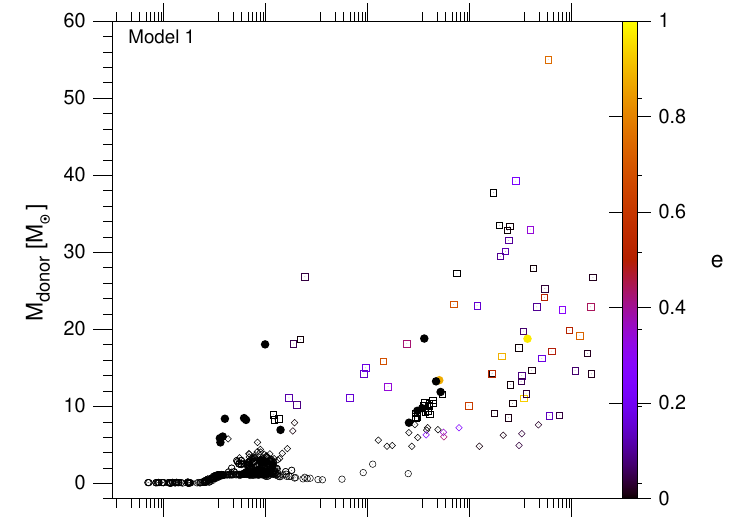}}
   \hspace{0.1cm}
     \resizebox{\hsize}{!}
   {\includegraphics{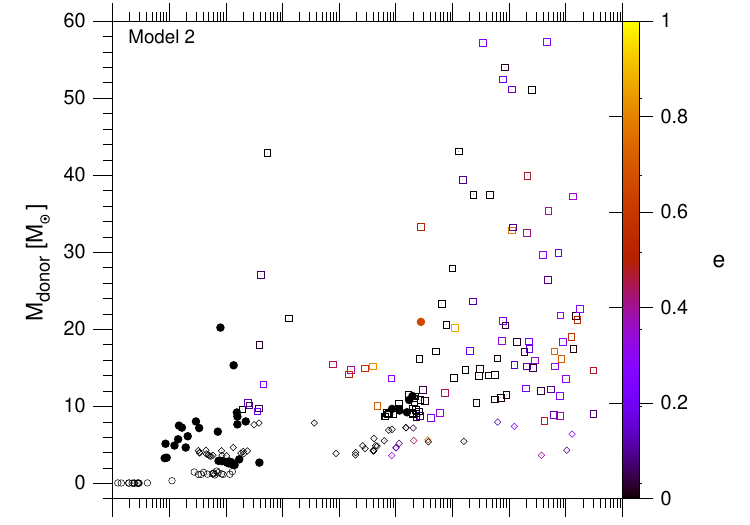}}
     \resizebox{\hsize}{!}
   {\includegraphics{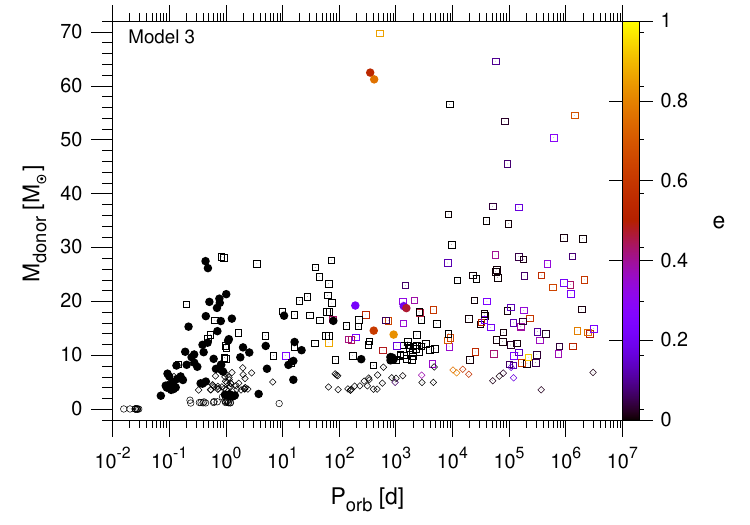}}
 \caption{The dependence between donor mass ($M_{\odot}$), orbital period (days) and eccentricity of XRBs for all 3 models. The eccentricity is color coded, the open symbols represent the population of XRBs described in Sec.~\ref{subsec:number_xrbs_group} and Sec.~\ref{subsec:number_xrbs_subgroups}: HMXBs are plotted as the open squares, IMXBs as open diamonds and LMXBs as open circles. The filled circles are XRBs that become merging DCOs generated in 10 runs for each model, as described in \ref{subsec:Discussion_mergingDCOs}. It means that the DCOs progenitors are 10 times over-represented with respect to other XRBs on the plots to show which part of the parameters space they occupy.}
  \label{Fig:XRBs_Ms_Porb_e}
\end{figure}

On Fig.~\ref{Fig:XRBs_Ms_Porb_e} we show the dependence between the donor mass, orbital period and eccentricity of XRBs in each model. The distribution of the systems along the vertical axis mirrors our definition of LMXBs, IMXBs and HMXBs: LMXBs (open circles) lay in the part of the diagrams where $M_{d}<3.0\,M_{\odot}$ for Model 1 and $M_{d}<3.5\,M_{\odot}$ for Model 2 and 3, all HMXBs (open squares) occupy the upper part of the diagram where $M_{d}>8.0\,M_{\odot}$ and IMXBs (open diamonds) lay in between. The maximum donor mass reaches $60\,M_{\odot}$ in Model 1 and 2 and increases to $\sim70\,M_{\odot}$ in Model 3. The orbital periods of all LMXBs in Model 2 and 3 are shorter than $\sim10$ days going down to an order of minutes, while in Model 1 there are 4 systems with orbital periods  $\sim100$ days. Both IMXBs and HMXBs span the wide range of $P_\mathrm{orb}$ from hours up to $\sim3\times10^6$ days. Systems with donors more massive than $3.5\,M_{\odot}$ and wider orbits ($P_{orb}>100$ d) may also be highly eccentric, however, most of them have eccentricities below $0.4$. All LMXBs have circular orbits what is an outcome of the combined effect of tidal circularization and common envelope phase that they undergo during their prior evolution. This figure is complemented by Fig.~\ref{Fig:progenit_XRBs_Porb_e} which illustrates the initial orbital periods and eccentricities of the binaries at ZAMS which are the progenitors of XRBs in our models.

\subsection{Merging DCOs}\label{subs:results_mDCOs}
\begin{figure*}
   \centering
   {\includegraphics[scale=0.7]{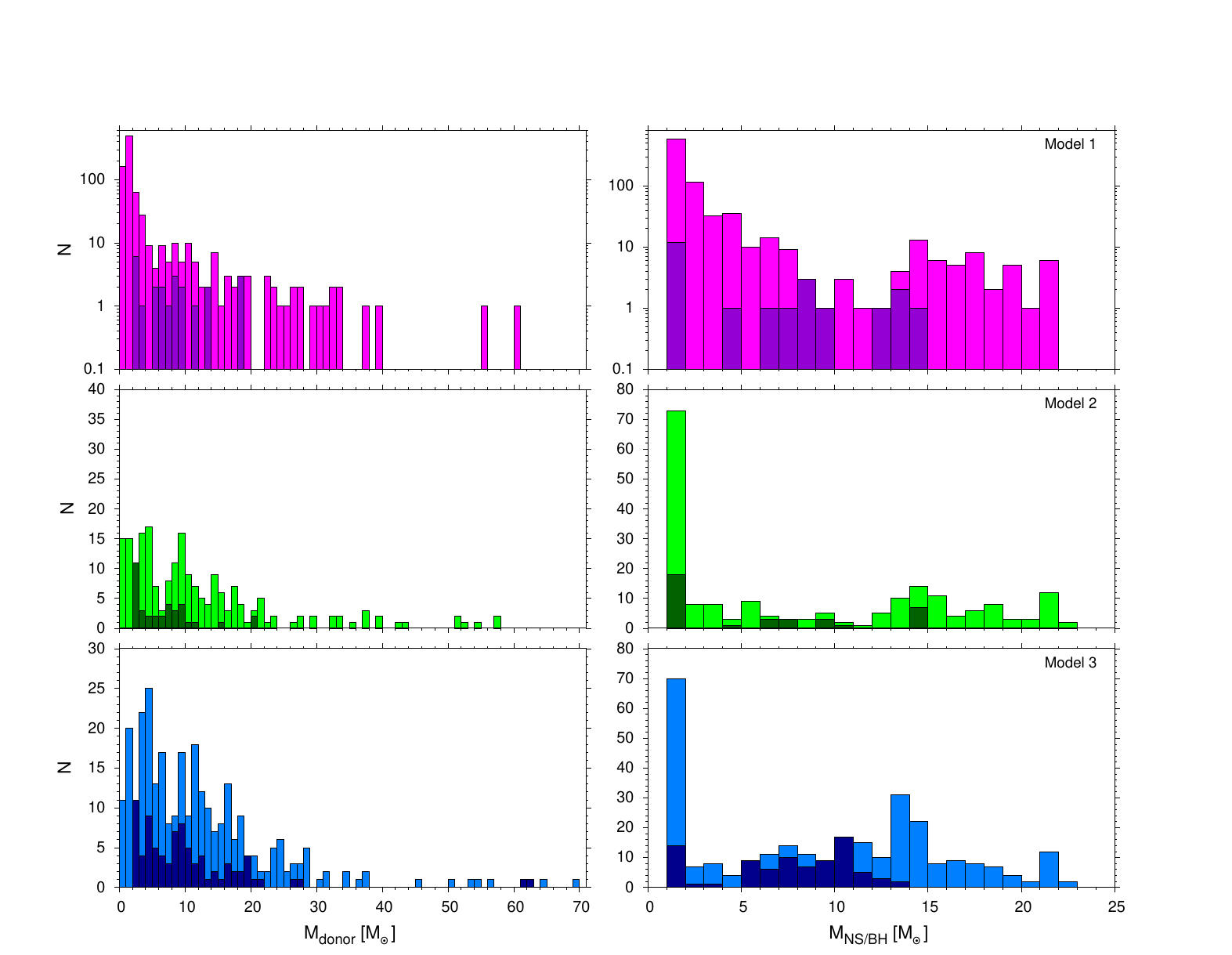}}
 \caption{The distributions of the donor (left column) and compact object (NS or BH) (right column) masses at the onset of XRB phase in the three models described in Sec.~\ref{Sec:results}. The darker colors on each panel correspond to the distributions of masses of XRBs that become merging DCOs generated in 10 runs for each model, as described in \ref{subsec:Discussion_mergingDCOs}. It means that the distributions for merging DCOs progenitors are 10 times over-estimated with respect to distributions of the other XRBs on the figure to show which part of the parameters space they occupy. }
  \label{Fig:XRBs_mass_distrb}
\end{figure*}

We follow the future evolution of three models: the reference Model~1, the best observations matching Model~2 and the model with alternative CE development criteria (Model~3) to see if any of them produces the potential LVK sources. For each model we iterated the calculations 10 times to see how the resulting number of merging DCOs changes due to the different subsets of the present-day XRBs drawn from the entire XRBs population by Monte Carlo method incorporated in {\fontfamily{cmtt}\selectfont StarTrack}. 
We conclude that the average number of merging DCOs per population of XRBs that can be observable in M83 is on average $\sim1.9$ for two standard models and $6.2$ for Model~3. Specifically in the model of choice (Model~2) the average number of merging DCOs is $2.0$ where BH-NSs mergers contribute the most ($0.8$ mergers on average) and NS-NS/BH-BH make equal contribution of $0.6$ mergers of each type (see Table~\ref{table:DCOs_types}). HMXBs are the progenitors of $1.2$ mergers and of them $0.5$ are BH-BH mergers, $0.2$ are BH-NS mergers and $0.5$ are NS-NS mergers. The remaining $0.8$ merging binaries originate from IMXBs ($0.6$ are BH-NS, $0.1$ are BH-BH and $0.1$ are NS-NS mergers).
In the consequence of the reduction of almost all LMXBs none of the merging DCO in Model~2 comes from LMXBs. However, as can be seen from the results of Model~1 (see Table~\ref{table:BestFit}) LMXBs may be a potential progenitors of merging NS-NS and BH-NS binaries. In Model 1 which contains much larger population of LMXBs than Model 2 and 3, we find on average $1.0$ NS-NS, $0.57$ BH-NS and $0.14$ BH-BH mergers. Both LMXBs and IMXBs account for $0.7$ of merging DCOs progenitors in Model~1, while HMXBs account for the remaining $0.3$ mergers.\\
The higher number of HMXBs and IMXBs in Model 3 than in standard Models 1 and 2 has its effect on the number and the type of merging DCOs. We find $3.22$ BH-BHs, $2.44$ BH-NSs and $0.56$ NS-NSs among an average of $6.22$ mergers in Model 3. All merging BH-BHs and NS-NSs have HMXBs as they progenitors, $0.88$ merging BH-NSs are the result of HMXBs evolution as well, but the majority of merging BH-NSs ($1.56$) goes through IMXBs phase in their prior evolution. In other words HMXBs and IMXBs are the progenitors of $4.7$ and $1.5$ merging DCOs in Model 3.\\
On Fig.~\ref{Fig:XRBs_mass_distrb} we show the donor mass and NS/BH mass distributions of all XRBs that become merging DCOs (dark-violet boxes for Model 1, dark-green boxes for Model 2 and dark-blue boxes for Model 3). We choose to present XRBs that become merging DCOs that we found in all 10 iterations for each model as mentioned at the beginning of this section to give a better view on the parameter space they take up. Therefore, their number is 10 times over-represented in comparison to XRBs that do not become merging DCOs showed on Fig.~\ref{Fig:XRBs_mass_distrb} in light colors.

   \begin{figure}
   \centering
   \resizebox{\hsize}{!}
   {\includegraphics{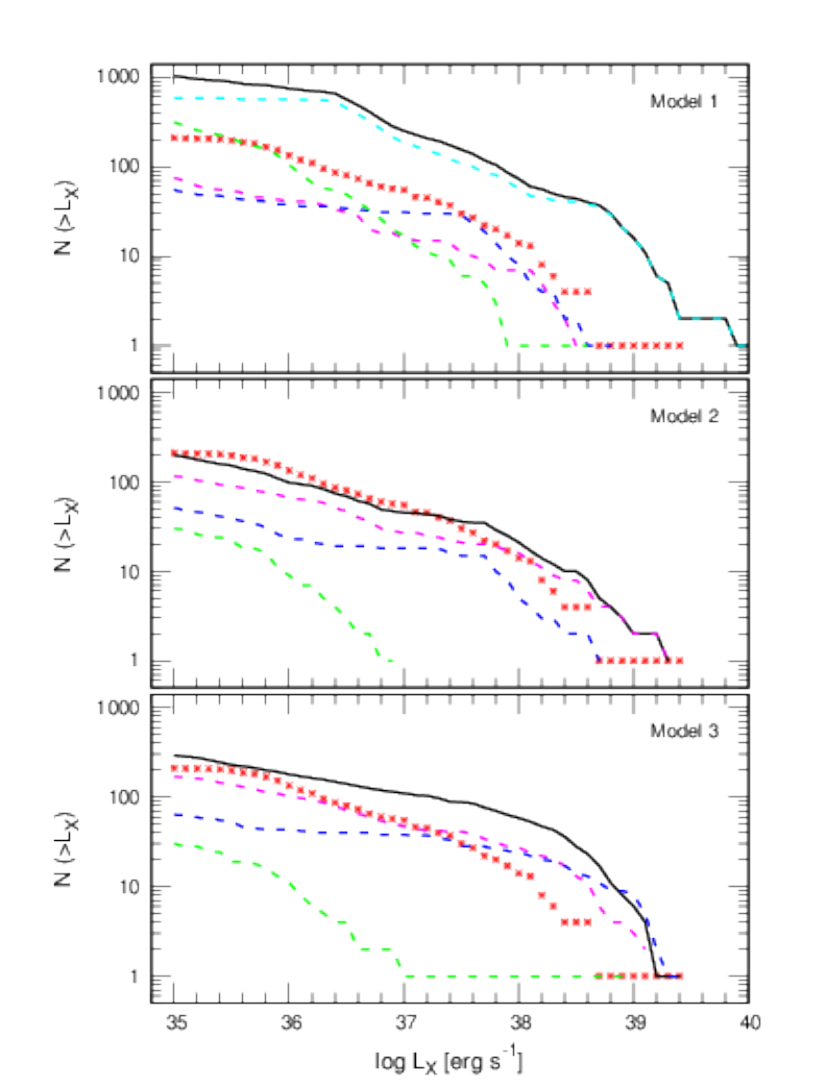}}
 \caption{The cumulative XLF of XRBs for M83: observations (red stars) and Model 1 (top panel), Model 2 (middle panel) and Model 3 (bottom panel). On the vertical axis is shown the number of XRBs which have luminosity $L_\mathrm{X}$ higher than the corresponding $L_\mathrm{X}$ value on the horizontal axis. The models are plotted with solid black lines (all XRBs) and subdivided into four subcategories shown with 
dashed lines: HMXBs (magenta), IMBXs (blue), persistent LMXBs (cyan), outbursting transient LMXBs (green) in Model 1 or randomly selected 30 LMXBs (green) in Model 2 and 3. Note that Model 2 is consistent with the observations in terms of total number of XRBs and the shape of XLF. Model 3 is close to the observations in terms of total number of XRBs but the slope of XLF is shallower in the range $\log L_\mathrm{X}\sim36.6-38.6\,\mathrm{erg\,s^{-1}} $  than in Model 1 and 2.}
  \label{xlf}
\end{figure}

\begin{table}
\caption{Observed and model numbers of XRBs for galaxy M83. Note that Model 2 is a good approximation 
of observations in terms of the total number of XRBs as well as its observed subclasses: LMXBs, 
IMXBs, and HMXBs. Model 3 is closer to observations in LMXBs and IMXBs numbers than Model 1 but is not as good as Model 1 due to the higher number of HMXBs. Numbers in parenthesis show the estimated number of XRBs which 
are predicted to end up as merging double compact objects (NS-NS/BH-NS/BH-BH). The numbers of outbursting transient LMXB (oT-LMXBs) and persistent LMXBs (P-LMXBs) in each model are also shown.}
\label{table:BestFit}      
\centering                        
%%\resizebox{\columnwidth}{!}{ 
\begin{tabular}{c | c c c c }        % centered columns (4 columns)
\hline\hline                 % inserts double horizontal lines
\noalign{\smallskip}
XRB & \multicolumn{4}{c}{Number of systems}\\    % table heading 
\noalign{\smallskip}
type & \textit{Observed}  & \multicolumn{3}{c}{Model} \\
 & & 1 & \textbf{2} & \textbf{3}\\
\noalign{\smallskip}
\hline                        % inserts single horizontal line
\noalign{\smallskip}
   LMXB & \textit{30}  & 689 (0.7) & \textbf{30} \, (0.03) & \textbf{30} (0.03) \\ 
  \footnotesize
  P-LMXB & \textit{?}  & \footnotesize 441 & \footnotesize \textbf{0} & \footnotesize \textbf{2} \\
  \footnotesize
   oT-LMXB & \textit{?}  & \footnotesize 248 & \footnotesize \footnotesize \textbf{30}  & \footnotesize \textbf{28} \\
   
   IMXB & \textit{64}  & 55 \, (0.7) & \textbf{51} \, (0.8) & \textbf{64} (1.5) \\
   HMXB & \textit{120} & 75 \,  (0.3) & \textbf{116}  (1.2) & 
  \textbf{169} (4.7) \\   
\hline    
\noalign{\smallskip}
Total & \textit{214} & 819 (1.70) & \textbf{197} (2.03) & \textbf{263} (6.23) \\
 \smallskip
\end{tabular}
\end{table}

\section{Discussion}\label{Sec:Discussion}
\subsection{The synthetic number of XRBs}\label{Subsec:Dicuss_XRBs_number}
The XLFs of the synthetic populations of XRBs that we generated with the standard models (i.e. with the standard CE development criteria) have the shapes that track the shape of the observational curve (see Fig.~\ref{xlf} for Models 1, 2). It means that for the models with the standard CE development criteria the XLF shape itself of this particular XRB population (in M83) is not a good indicator of the underlying XRBs population: the population that is dominated by LMXBs (689 LMXBs in the total of 819 XRBs in Model 1)  has almost the same shape as the population dominated by HMXBs (116 HMXBs in 
total of 197 XRBs in Model 2).\\
However, the shape of the XLF generated with the revised CE development criteria differs from observational, Model 1 and Model 2 XLFs: its slope is less steep up to the luminosity $L_\mathrm{X}\sim4.0\times10^{38}\,\mathrm{erg\,s^{-1}}$. What can be noticed is a significant difference in the shape of IMXBs XLF (blue dashed lines on Fig.~\ref{xlf}) between Model 3 and other two models:  IMXBs XLF in Model 3 does not drop at luminosity $\sim6.0\times10^{37}\,\mathrm{erg\,s^{-1}}$ but it extends further up to $L_\mathrm{X}\sim1.0\times10^{39}\,\mathrm{erg\,s^{-1}}$.\\
The higher number of luminous IMXBs in Model 3 is a result of the adopted alternative CE development criteria that allow the systems to evolve through the stable RLOF evolutionary channel. The binaries that would merge during CE phase initiated by the expanding more massive star evolving in Hertzsprung gap in the case of standard models, go through the stable RLOF instead and after the first SN they form IMXBs with higher donor and accretor masses and higher luminosities than IMXBs formed through CE channel, i.e. there are formed more IMXBs with donors close to the upper donor mass limit.\\
In the case of HMXBs their XLFs shapes are alike for all three models. The evolution through the stable RLOF channel has an effect in the shift of HMXBs XLF towards higher numbers on y-axis. The shape of HMXBs XLF in Model 3 is not as much affected  by the CE development criteria as for IMXBs XLF because the criteria affect HMXBs in their whole donor mass range.\\

Our main modeling issue was to match the numbers of XRBs observed in M83. Typical model numbers that we find are much too large for LMXBs and somewhat too low for IMXBs and HMXBs (see Model 1 in Table~\ref{table:BestFit}). 
We checked that changing $Z$ within the range observed in M83 has not much influence on the synthetic XRBs population. For $Z=0.01,0.02,0.025,0.03$ the total number of systems differs at most by a factor of $1.3$.\\
Apart from the CE development criteria the number of HMXBs in our simulations is influenced the most by the slope of IMF for massive stars and SFR, within the small set of parameters that we varied. The uncertainty of the IMF slope for the high-mass stars, is: $\alpha_3=-2.3\pm0.7$ \citep{Kroupa01}. Changing $\alpha_3=-2.3$ (Model 1: $75$ HMXBs and $55$ IMXBs) to $\alpha_3=-1.9$ generates more massive stars and thus 
more HMXBs ans IMXBs ($227$ HMXBs and $225$ IMXBs). It is apparent that just with change of IMF slope within its 
uncertainity we can easily match the number of observed HMXBs ($120$) and IMXBs ($64$) in M83 . \\
 As mentioned in Sec.~\ref{Sec:Method} we tested 4 values of SFR ($1.5,\,2.0,\,2.5,\,3.5\, \mathrm{M}_{\odot}\,\mathrm{yr}^{-1}$). Only the models with  $\mathrm{SFR}=3.5\,\mathrm{M}_{\odot}\,\mathrm{yr}^{-1}$, standard IMF slope and $Z=0.01$ (e.g. Model 2) produce the number of HMXBs ($116$) approaching the observations ($120$). The lower is SFR the less HMXBs are formed in simulations, e.g. for the model with $\mathrm{SFR}=2.5\,\mathrm{M}_{\odot}\,\mathrm{yr}^{-1}$ and $Z=0.01$ (Model 1) their number drops to $75$.\\
From discussion above it is apparent that finely tuning either of the two parameters (SFR or $\alpha_3$) leads to the desired result considering HMXBs. However, there are other observations that motivate the specific choice of SFR that adjusts the number of HMXBs. \citet{Grimm03} inferred the relation between SFR and the total number $N$ of HMXBs  with luminosities $L_\mathrm{X}>2.0\times10^{38}\,\mathrm{erg\,s^{-1}}$ from the observations of $14$ local galaxies: 
\begin{equation}\label{eq:SFR_Grimm}
    \mathrm{SFR}(\mathrm{M_{\odot}\,yr^{-1}})=\frac{N(L>2\times10^{38}\,\mathrm{erg\,s^{-1}})}{2.9}
\end{equation}
This relation predicts that for  $\mathrm{SFR}=3.5\,\mathrm{M}_{\odot}\,\mathrm{yr}^{-1}$ the number of luminous HMXBs should be $N\sim10$. Our model with $\mathrm{SFR}=3.5\,\mathrm{M}_{\odot}\,\mathrm{yr}^{-1}$, $\alpha_3=-2.3$ and $Z=0.01$  (Model 2) gives $9$ such systems. Equation \ref{eq:SFR_Grimm}  predicts only $N\sim7$ for $\mathrm{SFR}=2.5\,\mathrm{M}_{\odot}\,\mathrm{yr}^{-1}$ while we find $17$ such binaries in model where $\mathrm{SFR}=2.5\,\mathrm{M}_{\odot}\,\mathrm{yr}^{-1}$, $\alpha_3=-1.9$ and $Z=0.01$.\\

We note that we did not consider transient IMXB/HMXBs with Be-star or Oe-star donors (Oe/Be-XRBs) in the models.
Some of our XRBs with normal O/B star donors (core H-burning) can be Oe/Be-XRBs, and if found in quiescence they 
should be removed from our model populations if they are in quiescence. The quiescence luminosity of these systems is $L_\mathrm{X}\leqslant1.0\times10^{33}\,\mathrm{erg\,s^{-1}}$ \citep[e.g.][]{Negueruela10}  which is below \textit{Chandra} detectability limit for M83. We make a rough estimation of how many HMBXs should be considered quiescent in our model of choice (Model~2) following the assumptions in \citet{Bel_Ziol_2009}: there are $10$ systems out of $116$ HMXBs that have companions burning H in their core, assuming only $25\%$ of them are Be-XRBs \citep[e.g.][]{McSwain05_BEstars} and that their duty cycle is $10-30\%$ \citep{Bel_Ziol_2009} at most $\sim1$ of them is quiescent Be-XRBs. Excluding it from our sample would have small influence on our conclusions, but it also implies that our number of HMBXs in a given model is an upper limit.\\
While the agreement between the observational and synthetic number of HMXBs in models with $\mathrm{SFR}=~3.5\,\mathrm{M}_{\odot}\,\mathrm{yr}^{-1}$, $\alpha_3=~-2.3$ and $Z=~0.01$ is satisfying, the number of IMXBs is still too high ($95$ compared to $64$ observed). This was the motivation to look closer at the mass limits adopted in the classification of XRBs into three groups (see Sec.~\ref{subs:X}). The mass limits taken from the comparison of the optical observations to evolutionary tracks in H21 are accurate to $1.0\,\mathrm{M}_{\odot}$ due to the fact that stellar evolution tracks are probed every $1.0\,\mathrm{M}_{\odot}$. Therefore, the stars of the mass below $3.5\,\mathrm{M}_{\odot}$ may be classified as  $3.0\,\mathrm{M}_{\odot}$ stars.
It appears that in the sample of $95$ IMXBs $44$ have the secondaries with the masses in the range $3.0\,\mathrm{M}_{\odot}<M_2<3.5\,\mathrm{M}_{\odot}$. Changing the condition for the maximum mass of the donor in LMXBs from $M_2<3.0\,\mathrm{M}_{\odot}$ to $M_2<3.5\,\mathrm{M}_{\odot}$ lowers the number of IMXBs to $51$ and brings it closer to the number derived from the observations.\\ 
 The main problem for us to solve is the striking excess of LMXBs produced in the simulations ($689$ in Model 1) in comparison to only $30$ LMXBs identified in M83. LMXBs are subdivided into two major categories; persistent LMXBs 
($441$ in Model 1) and transient LMXBs in outburst ($248$ in Model 1; as quiescent transients are not bright enough 
to make \textit{Chandra} detection threshold).
The division between transient and persistent LMXBs strongly depends on the disk stability limits determined by 
the disk instability model (see Sec.~\ref{subs:transients}). Although the adopted model is successful in prediction 
of the transient or persistent nature of the observed LMXBs \citep{Lasota08,Coriat12} the ``model calibration'' 
(see parameter $C_\mathrm{irr}$ in Eq.~\ref{eq:mcrit_He} and Eq.~\ref{eq:mcrit_solar}) is still highly uncertain. 
One could think of shifting as many P-LMXBs into T-LMXB category, and then removing them by putting some of them in 
quiescence (too low luminosity to make \textit{Chandra} threshold for the observation dataset that we employ). However, we have 
already adopted the lower limit on $C_\mathrm{irr}=0.001$ \citep{Tetarenko18}  minimizing the number of P-LMXBs. 
Higher $C_\mathrm{irr}$ leads to the decrease of the critical accretion rate above which LMXBs become persistent 
X-ray sources what results in more P-LMXBs in our synthetic populations.\\
It was noted that for LMXBs with very low donor to accretor mass ratio ($q\leq0.02$) the disk stability criteria 
presented in Sec.~\ref{subs:transients} may not work due to tidal forces operating on the accretion flow 
\citep{Yungelson06}. However, as pointed out by \citet{Lasota08} the maximum outburst luminosities inferred from the adopted disk model (Eq.~\ref{eq:mcrit_He}) for helium ultracompact X-ray binaries are consistent with 
observations what indicates that the adopted model works at least for some very low mass ratio LMXBs.\\
 In the case of T-LMXBs the number of systems in outburst is defined by the adopted duty cycle: $2.5\%$ of time spent 
in outburst \citet{Yan15}. Duty cycle is subject to large uncertainties and spans the range $0.9-7\%$ (see also 
\ref{subs:transients}). By lowering duty cycle from $2.5\%$ to $0.9\%$ we can lower the fraction of the ``visible'' 
T-LMXBs in the outburst in our models (e.g., from $248$ to $89$ for Model~1). But this hardly improves the issue.
To agree with observations we remove ("by hand") all but 30 randomly chosen LMXBs from our synthetic 
populations (Model~2 and 3). 
By choosing randomly 30 LMXBs we admit that we are not able to reproduce even such basic property of LMXBs population 
as the number of LMXBs sources observed in well known nearby galaxy. This failure must lead to future study of 
evolutionary processes that are important factors in the formation of LMXBs (mass transfer events in close binaries,
natal kicks received by compact objects, magnetic braking among other uncertain evolutionary factors). 

 \begin{figure}
   \centering
  \resizebox{\hsize}{!}
   {\includegraphics{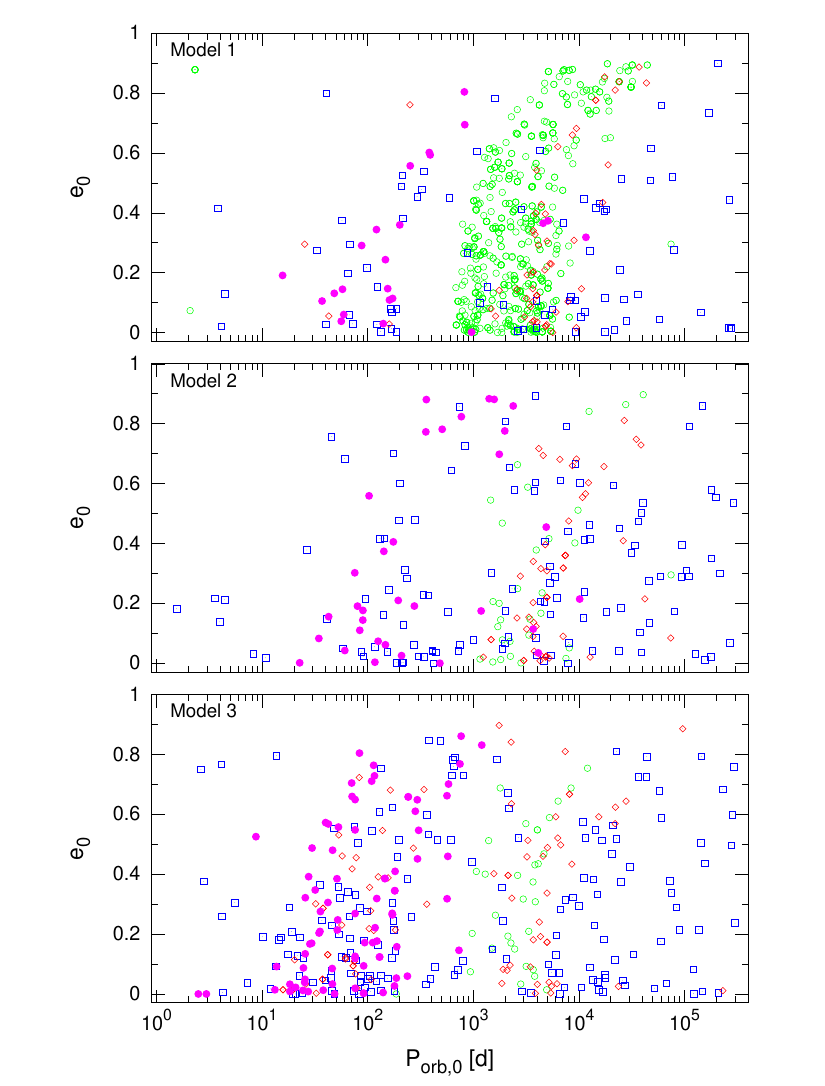}}
 \caption{The dependence between initial orbital period (days) and eccentricity of the XRBs progenitors in Model 1 (top), Model 2 (middle) and Model 3 (bottom ). The HMXBs progenitors are marked as blue open squares, IMXBs are red open diamonds and LMXBs are green open circles. The filled magenta circles are the progenitors of XRBs that become merging DCOs generated in 10 runs for each model, as described in \ref{subsec:Discussion_mergingDCOs}.  It means that the DCOs progenitors are 10 times over-represented with respect to other XRBs on the plots to show which part of the parameters space they occupy.}
  \label{Fig:progenit_XRBs_Porb_e}
\end{figure}

\subsection{XRBs characteristic}\label{subsec:Discuss_XRB_character}
The XRBs in the synthetic population form from the binaries that have wide range of initial orbital periods and eccentricities as can be seen on Fig.~\ref{Fig:progenit_XRBs_Porb_e}. The XRBs-progenitors are widely scattered in the parameter space in all 3 models. The orbital parameters of formed XRBs are shaped by:
\begin{enumerate}[label=(\roman*)]
    \item {the assumed criteria of CE development}
    \item{the mass loss of the massive stars due to their stellar winds which widens the orbit and changes the mass ratio of the binary}
    \item{the natal kick that the system encounters during SN explosion which randomly changes the eccentricity and orbital separation of the binary}
    \item{the time period during which the tides between the binary components may act to circularize the orbit}
\end{enumerate}
The interplay between those factors results in population of XRBs plotted in open symbols on Fig.~\ref{Fig:XRBs_Ms_Porb_e} for all 3 models. \\
The impact of different CE development criteria applied in Model 1/2 and Model 3 on HMXBs and IMXBs populations can be seen when one compares Fig.~\ref{Fig:progenit_XRBs_Porb_e} and Fig.~\ref{Fig:XRBs_Ms_Porb_e}. HMXBs have $P_\mathrm{orb}>100$ days in Model 1/2 while they are more evenly distributed on the orbital periods values in Model 3. IMXB-progenitors are concentrated around $P_\mathrm{orb}\sim$ few $10^3$ days while in Model 3 they can have also $P_\mathrm{orb}\sim10-10^2$ days. This is because the systems that would undergo CE phase and merge in Model 1(2) go through the stable mass transfer and survive as HMXBs/IMXBs in Model 3. LMXBs are insensitive to the revised CE development criteria from the definition of the latter given in Sec.~\ref{Sec:Method}. Their progenitors have to have orbital periods of order of $10^3$ days to survive the subsequent evolution. They form LMXBs with orbital periods shorter than 2 days. The LMXBs-progenitors have all range of eccentricities but the circularization of an orbit which is an interplay between the tides and the Roche lobe overflow mass transfer which takes place in the LMXB-phase causes that all synthetic LMXBs have $e=0$.

\subsection{Merging DCOs}\label{subsec:Discussion_mergingDCOs}

To better present the characteristic of XRBs that form merging DCOs we present separately all such XRBs from 10 different subsets of the present-day XRBs in M83 drawn from the entire XRBs population as dark histograms on Fig.~\ref{Fig:XRBs_mass_distrb} and filled circles on Fig.~\ref{Fig:XRBs_Ms_Porb_e} and on Fig.~\ref{Fig:progenit_XRBs_Porb_e} (we do not present the whole present-day XRBs population from all 10 runs for clarity of Fig.~\ref{Fig:XRBs_Ms_Porb_e}). It means that the progenitors of merging DCOs are 10 times over-represented on all those figures. Although it may be visually misleading our purpose is to show which part of the parameter space is occupied by such systems. The densely populated part of the top panel of Fig.~\ref{Fig:XRBs_Ms_Porb_e} is zoomed on Fig.~\ref{Fig:Model1_zoom} in \ref{App:tables}.\\
As can be seen on Fig.~\ref{Fig:progenit_XRBs_Porb_e} most of the progenitors of XRBs that form merging DCOs (magenta filled circles) span the whole range of the initial eccentricities but the values of their initial orbital periods $P_\mathrm{orb,0}$ are tightly connected to the criteria for the common envelope development that are applied in their subsequent evolution: most of them have $P_\mathrm{orb,0}$ in the range between tens to hundreds of days in Model 3 while there are the progenitors that have much longer periods of order of few thousands of days in Model 1 and 2. The orbit of the merging DCOs progenitors cannot be initially too wide in Model 3 because most of the merging DCOs progenitors evolve through the stable mass transfer channel and the shrinkage of their orbit will not be as dramatic as during CE phase, which is a dominant evolution channel for merging DCOs in Model 1 and 2.
All merging DCOs in our 10 samples of each model form from XRBs with orbital periods lower than $\sim2\times10^3$ days. In Model 1 and 2 the merging DCOs progenitors group at the two orbital period ranges: the short-period range from $\sim2.4$ hours to a few days and the long period range around $10^3$ days, there are no merging DCOs progenitors with orbital periods in between. In the long orbital period group there are only XRBs with high-mass donors, i.e. these are the systems that become XRBs due to the wind accretion before they undergo the common envelope phase. All systems in the short-group become XRBs after CE. In the model with the revised CE development criteria (Model 3) the XRBs resulting in merging DCOs are more evenly distributed, however still most of them have $P_\mathrm{orb}\sim2$ hr to $\sim2$ days. The lack of the gap in the orbital period distribution in Model 3 illustrates the characteristic feature of this model: there exist systems which undergo the stable mass transfer rate shrinking their orbit gradually in the regime not available for the merging DCOs progenitors in Model 1 and 2.
\\
There are only a few systems that we find in the population of merging DCOs-to-be XRBs in our 10 runs that have eccentric orbits: 2 in Model 1, 1 in Model 2 and 7 in Model 3. All these systems are HMXBs in which tidal forces have not yet circularized the orbit at the time that they become wind-fed XRBs.

The merging DCOs constitute $\sim1\%$ (Model 1 and 2) to $\sim2\%$ (Model 3) of the intrinsic XRBs population in M83. The dominant merging DCO type depends on the model: it is NS-NS in Model 1, BH-NS in Model 2 and BH-BH in Model 3 (see Table~\ref{table:DCOs_types}). At this point we can also compare the fraction of HMXBs that will merge as BH-BH in each model with the expectations inferred by \citet{Liotine2023}. They estimate that $\sim0.6\%$ of detectable HMXBs will merge as BH-BH withing the Hubble time. From our synthetic population of HMXBs detectable in M83 we obtain that $\sim0.22\%$, $\sim0.43\%$ and $\sim1.89\%$ of such systems will merge as BH-BH in Model 1, 2 and 3 respectively. The number from Model 2 is the closest to the estimations of \citet{Liotine2023} what adds up to the arguments in favor of Model 2 as being the model of choice.\\
In the models in which we reduce the number of LMXBs to $30$ systems (Model 2 and 3) the most probable progenitors of merging DCOs are HMXBs but in the model in which P-LMXBs and outbursting T-LMXBs are not arbitrarily removed (Model 1) the population of LMXBs contribute to the number of merging DCOs to the same extend as population of IMXBs (see Table~\ref{table:BestFit}. However, the probability that LMXBs will evolve into a merging NS-NS is much lower than for IMXBs or HMXBs in Model 1: only 1 in 1000 LMXBs may result in merging DCO comparing to 12 in 1000 IMXBs and 4 to 1000 HMXBs. Assuming the same effectiveness of merging DCOs production from LMXBs in Model 2 and 3 we estimate that from 30 LMXBs we can expect $\sim0.03$ mergers in those models.\\
Regarding the uncertainty that we have about the LMXBs evolution we can not exclude the possibility that $\sim1$ LMXB forming merging DCO does not appear in the population of LMXBs in Model 2 and 3 only due to our artificial selection of 30 systems. We have no means to tell if the population of the observable LMXBs in M83 contains or does not contain the system that will merge as NS-NS in the future. To assess the consequences of our approach to LMXBs in Model 2 and 3 let's assume that 30 LMXBs in Model 2/3 drawn in 10 runs give  $0.7$ (at most) merging DCOs like in Model 1. This assumption is equivalent to the situation when 30 LMXBs would have been drawn from the part of LMXBs parameter space which is the most probable to result in merging DCO. This changes the expected number of merging DCOs by $\sim24.8\%$ (from $2.03$ to $2.70$) in Model 2 and by $\sim9.7\%$ in Model 3 (from $6.23$ to $6.90$). The change of the number of merging DCOs is not negligible, especially in Model 2, but it does not influence our conclusions.\\

In all three models merging BH-BHs come from HMXBs with accreting BH (BH-HMXBs). The typical evolution leading to merging BH-BH starts from the binary composed of two massive stars. When the initially more massive star (primary) ends to burn hydrogen in its core its expansion leads to thermal time-scale mass transfer (TTMT) on the less massive star (secondary) which is still on the main sequence. The primary loses almost its entire hydrogen envelope in the process what inhibits its rapid expansion. The mass transfer ceases when the radius of the primary Roche lobe becomes larger than the radius of the primary itself. As the evolution of the donor proceeds it becomes the naked He-star which forms the first BH and the system becomes wide  BH-MS binary. Soon after the secondary evolves into the core He-burning star (CHeB) its radius surpasses its Roche lobe radius. Depending on the CE development criteria adopted in the model the future evolution of BH-CHeB binary may go through two different channels leading to the final merger: the CE channel (the standard criteria in Model~1 and 2 \citet{B08}) or the stable RLOF channel (the alternative criteria in Model~3, \citet{Olejak21_CE}).\\
In the CE channel BH-CHeB goes through the common envelope and the donor becomes the naked He-star. The system becomes HMXBs which lives for $\sim 1.0$ Myr. In subsequent evolution the system survives the second SN explosion becoming tight BH-BH binary which merges within the Hubble time. In the stable RLFO channel, after the first BH is formed, the second star leaves the main sequence, expands and initiates TTMT onto BH but here the alternative CE development criteria allow the system to avoid the formation of CE. During stable TTMT phase all systems which form merging BH-BHs in Model~3 have heavy donor star transferring mass onto the light BH ($q\sim4-5$) and their orbits shrink significantly (about 2 orders of magnitude). The mass transfer ceases when the donor becomes the naked He-star and the subsequent evolution of the binary follows the same path as in Model~1 and 2 leading to the formation of merging BH-BH.\\
Most merging BH-NS binaries in all three models come from IMXBs with accreting BH (BH-IMXBs) which follow the CE evolution channel similarly to BH-HMXBs in Model 1 and 2 as described above. The alternative CE development criteria are not fulfilled here because of the donor ZAMS mass is below $18\mathrm{M_\odot}$ in these systems (see Sec.~\ref{Sec:Method}). After CE phase the donor becomes evolved helium star, the system mass ratio is now low enough to allow for the stable mass transfer onto BH and the system appears as IMXBs. Further evolution results in the second SN and NS formation. After few hundreds of Myrs from its formation the binary consisting of NS and BH merges.\\
If we exclude almost all LMXBs from the population 
(Model~2 and 3) all merging NS-NS binaries form from HMXBs when NS is a compact object (NS-HMXB). The evolutionary paths of NS-HMXBs are similar to these of BH-HMXBs in the same models. The difference is that when the binary goes through the CE stage the donor is already core He-burning H-rich star which subsequently becomes the evolved helium star. When CE phase ends the system is close enough for the donor to start the mass transfer onto NS. While transferring mass the donor proceeds onto the giant branch of He-stars. If the system survives the subsequent SN explosion and its orbital separation is sufficiently small it ends up as merging NS-NS. \\
In the model where all outbursting T-LMXBs and P-LMXBs are included (Model~1) the population of LMXBs may have a non-negligible contribution to the number of the progenitors of the merging NS-NS (see Table~\ref{table:BestFit} and discussion above in this section). Such systems are possible to form from LMXBs when the donor to the first-born NS is a helium star with the mass $2.6\leq M\leq 6.7\, \mathrm{M_\odot}$ which subsequently evolves to become the second NS on the close orbit \citep{Belczynski01_NSNS,Dewi02}. \\
In the typical evolution of LMXBs into the merging NS-NS in Model 1 the more massive main sequence star evolves into the Hertzsprung gap star where its rapid expansion leads to TTMT. When the mass ratio reverses and the initially more massive primary loses almost $50\%$ of its mass the mass transfer changes from TTMT to nuclear RLOF during which the mass in transferred from less massive to more massive binary component. When the mass transfer stops the primary is a red giant star. Finally it becomes the evolved helium star which explodes as SN leaving a NS behind. The evolution of the secondary proceeds. It becomes a red giant which at some point initiates the common envelope phase resulting in NS-naked He-star binary on an orbit of a few solar radii. Then the expanding He-star initiates the second episode of TTMT. While transferring mass the secondary evolves to the helium giant branch star which turns into NS after second SN event. The NS-NS binary which forms mergers within few hundreds Myr.\\
The evolutionary paths of the systems leading to merging DCOs in our simulations are given in Tables~\ref{table:EvolPaths_Model12} and \ref{table:EvolPaths_Model3} and the schematic pictures of the typical evolution of merging BH-BH from HMXB, NS-BH from IMXB and NS-NS from LMXB in Model 1 are included in Appendix.\\

To understand the scarcity of the merging DCOs in the synthetic population we follow the evolutionary paths of HMXBs, IMXBs and LMXBs which do not form merging DCOs. We give the numbers for the model of choice (Model~2) but the described evolution is valid in all 3 models.\\
There are 4 typical evolutionary paths which prevent the formation of merging DCO from HMXB. Among all HMXBs which do not end up as merging DCOs:
\begin{enumerate}[label=\alph*)]
\item more than half ($\sim56.7\%$) are formed as wide binaries with a core-helium hydrogen-rich giant donors where the accretion takes place through the stellar wind. Wide orbit means that the binding energy of the binary is small and as a result most of these systems are disrupted after the supernova explosion of the secondary.
\item $\sim24.3\%$ remain bound with two compact objects orbiting each other but their separation (of order of a few thousands $\mathrm{R}_\odot$) is too large for the merger to appear within the Hubble time.
\item $\sim11.7\%$ (only the ones with NS) go through CE and tighten their orbits to tens of $\mathrm{R}_\odot$ but the natal kicks imparted to the newly formed NS after the second SN are strong enough to disrupt these systems.
\item the remaining $\sim7.3\%$ form HMXBs with a main-sequence donor after the first SN and with the orbital periods small enough that they merge when the donor becomes low-mass He-giant and initiates the common envelope phase.
\end{enumerate}

In the population of IMXBs which do not merge as DCO at their final stage of evolution almost all BH-IMXBs end up as wide, non-interacting binaries where the companion star is a CO white dwarf (WD) and just a few percent of BH-IMXBs are disrupted after the second SN which leads to the birth of the second BH. The final fate of NS-IMXBs is more diversified: 
{\em (i)} the majority becomes close binaries but with merging time well beyond Hubble time, among them about one half are NS-NS binaries and second half are NS-WDs (CO or ONe WD); {\em (ii)} some form NS-hybrid WD systems which eventually merge and {\em (iii)} a few of the systems fall apart after the creation of the second NS.\\
In Model~1 almost all LMXBs which do not form merging DCOs end their lives either as NS-NS/WD-NS which do not merge in Hubble time, or as the binaries where the companion star merges with NS/BH before it becomes a compact object.\\

In this work we consider only the isolated binary evolution scenario. Among alternative scenarios for the evolution of the progenitors of merging DCOs the most widely considered is the scenario of the dynamical interactions in dense stellar clusters. The question may arise if the XRBs from the dynamical interactions could have the contribution to the observed XRBs population in M83. However, the numerical simulations that investigate the possible population of dynamically formed XRBs in Milky Way show that it is unlikely that they significantly contribute to the observed population of LMXBs \citep{Kremer18}. HMXBs follow the regions of higher SFR and there is some evidence that they may be associated with young stellar clusters \citep{Kaaret04,Garofali12} that can be found in the same regions. But though HMXBs may be born in such clusters the simulations done for HMXBs in NGC~4449 show that it is very unlikely that they were formed dynamically \citep{Garofali12}.  We assume that these conclusions also apply to M83.

\section{Conclusions}\label{Sec:concl}

We performed numerical simulations on nearby galaxy M83 to reproduce its X-ray binary populations and 
to predict its capability to form merging double compact objects (sources of gravitational-waves) out 
of these X-ray binaries. We find that:

\begin{enumerate}
    
\item{We can match the shape of the observed X-ray luminosity function, the number of XRBs and their 
specific subcategories (LMXBs, IMXBs and HMXBs) for the evolutionary channel that allows for effective 
formation of merging double compact objects through common envelope evolution (Model 2).
However, this match can be obtained only with artificial drastic reduction of LMXBs from our synthetic 
population. This shows that the physics involved in the formation of LMXBs is not well understood and 
requires further in-depth investigation. Also, it is argued that our standard CE development criteria 
may be too optimistic in the light of the recent detailed evolutionary studies.}     

\item{The evolutionary model in which majority of merging double compact object formation occurs with 
help of stable RLOF (without assistance of common envelope; Model 3) does not provide good match to the 
observed XLF shape. However, we note that this model is rather extreme and restrictive in its assumptions 
on CE development criteria. Possibly, a more realistic model that can form merging double compact objects 
through CE and stable RLOF (something between Model 2 and 3) in more balanced proportions is required in 
future studies.} 

\item{Independent of our adopted evolutionary scenario only $\sim 1-2\%$ of M83 XRBs will form merging 
double compact objects. This comes from the fact that evolution terminates X-ray binaries in binary 
component interactions, binary disruptions at second supernova or leads to the formation of wide 
(non-merging) double compact objects.} 

\end{enumerate}

Note that our conclusions are not a general statement about XRBs and their merging DCO production efficiency, 
but they  apply only to this specific galaxy with rather high metallicity. Yet, most of local galaxies for 
which  X-ray observations exist have typically high metallicity. Despite the fact that the connection between 
local XRBs and merging DCOs is rather weak, studies of XRBs populations may help constrain uncertain 
evolutionary physics that is involved in formation of LIGO/Virgo/KAGRA sources.
\begin{acknowledgements}
      We thank the anonymous referee for their useful suggestions and comments that helped to improve this paper.
      This work was supported by the Polish National Science Center (NCN) grant Maestro (2018/30/A/ST9/00050). I.K. would like to thank Aleksandra Olejak, Amedeo Romagnolo and Alex Gormaz-Matamala for valuable discussions and comments.    
\end{acknowledgements}

\bibliographystyle{aa} % style aa.bst
\bibliography{M83.bib}

\begin{appendix}\label{Sec:App1}

\section{Additional tables and figures}\label{App:tables}
\begin{figure}
   \centering
   \resizebox{\hsize}{!}
   {\includegraphics{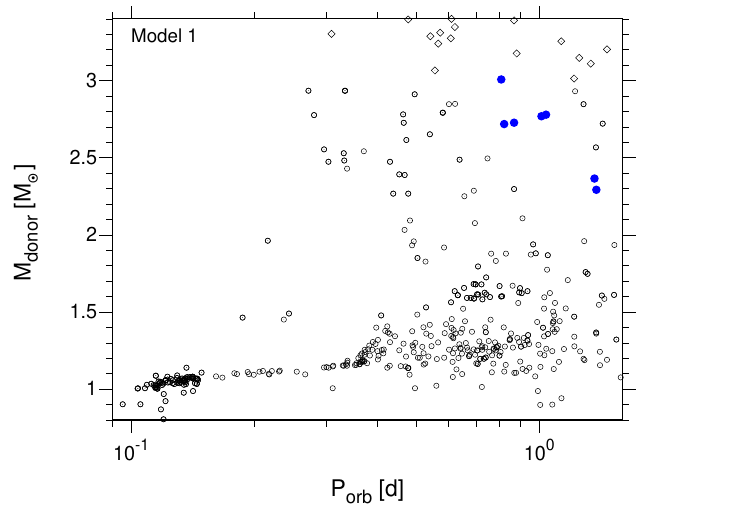}}
 \caption{The zoomed part of the top panel (Model 1) of Fig.~\ref{Fig:XRBs_Ms_Porb_e} (described in Sec.~\ref{subsec:Discuss_XRB_character} and Sec.~\ref{subsec:Discussion_mergingDCOs}) where the points cluster the most. Blue filled circles correspond to XRBs that form merging DCOs and open symbols show the rest of XRBs synthetic population in Model 1 (see Sec.~\ref{Sec:results}). The figure shows that there are no XRBs that form merging DCOs with $M_\mathrm{donor}<2.0\,M_{\odot}$ and $P_\mathrm{orb}<2.0$ days. We note that the merging DCOs progenitors are 10 times over-represented with respect to other XRBs on the plot to show which part of the parameters space they occupy.}
  \label{Fig:Model1_zoom}
\end{figure}

Here we present two additional tables and one figure which complement the main text. In Table~\ref{Table:Models_factors} we summarize the values of the reduction factors calculated for each model described in Sec.~\ref{Sec:predicted_number_factors}. In Table~\ref{table:DCOs_types} we distinguish the contribution of  different types of merging DCOs in total number of merging DCOs from XRBs in Model 1, 2 and 3 given in Table~\ref{table:BestFit} described in main text in Sec.~\ref{subs:results_mDCOs}.\\
Fig.~\ref{Fig:Model1_zoom} gives the better insight in the most clustered region on left bottom part of the top panel (Model 1) of Fig.~\ref{Fig:XRBs_Ms_Porb_e} described in Sec.~\ref{subsec:Discuss_XRB_character} and Sec.~\ref{subsec:Discussion_mergingDCOs}. We show here that there are no merging DCOs progenitors with $M_\mathrm{donor}<2.0\,M_{\odot}$ and $P_\mathrm{orb}$ days.

\begin{table}[ht!]
\caption{The reduction factors described in Sec.~\ref{Sec:predicted_number_factors} for each of the three models.}        % title of Table
\label{Table:Models_factors}      % is used to refer this table in the text
\centering                       % used for centering table
%\resizebox{\columnwidth}{!}{ 
\begin{tabular}{c |c c c c }        % centered columns (4 columns)
\hline\hline                 % inserts double horizontal lines
\noalign{\smallskip}
Model & $f_{\mathrm{L_X}}$ & $f_{\mathrm{vis}}$ & $f_{\mathrm{P}}$ &  $f_{\mathrm{DC}}$ \\
\noalign{\smallskip}
\hline                     % inserts single horizontal line
\noalign{\smallskip}
1 & 0.274 & 0.752 & 0.042 & 0.025\\      
2 & 0.271 & 0.752 & 0.043 & 0.025\\
3 & 0.275 & 0.752 & 0.043 & 0.025 \\   
   \hline  
\end{tabular}
\end{table}

\begin{table}[ht!]
\caption{The average number of merging DCOs of a given type in three models described in Sec.~\ref{Sec:results}.}
\label{table:DCOs_types}      
\centering                        
%%\resizebox{\columnwidth}{!}{ 
\begin{tabular}{c | c c c }        % centered columns (4 columns)
\hline\hline                 % inserts double horizontal lines
\noalign{\smallskip}
DCO & \multicolumn{3}{c}{Model}\\    % table heading 
\noalign{\smallskip}
type & 1 & 2 & 3 \\
\noalign{\smallskip}
\hline                        % inserts single horizontal line
\noalign{\smallskip}
   NS-NS & 1.00 & 0.60 & 0.56 \\      
   BH-NS & 0.57 & 0.80 & 2.44 \\
   BH-BH & 0.14 & 0.60 & 3.22 \\   
\hline    
\end{tabular}
\end{table}

\section{Model 4 - the model with rapid SN engine.}
We check how the model with SN mechanism that produces the mass gap in compact objects mass distribution (rapid SN engine \citep{Fryer12}) changes the results comparing to the three models described in the main text. There is a significant increase in the number of XRBs in Model 4 comparing to Models 1-3. The quick explanation is that the BHs produced in the rapid SN model have masses above the mass gap so they are on average more massive than BHs in the delayed SN model. Hence more BHs experience weaker natal kicks then in the rapid SN model which allows more systems to survive. For detailed discussion of the impact of SN engine on subsequent evolution of the binaries with NS/BHs see \citet{Olejak22_SNengine}. We conclude that apparently, the rapid model produces even larger tension with the observations.

\begin{table}
\caption{Observed and model numbers of XRBs for galaxy M83. The model here is the variation of Model 1 with rapid instead of delayed SN engine (Model 4). As can be seen from numbers Model 4 produces even larger tension with observation than Model 1. Numbers in parenthesis show the estimated number of XRBs which 
are predicted to end up as merging DCOs (NS-NS/BH-NS/BH-BH). The numbers of outbursting transient LMXB (oT-LMXBs) and persistent LMXBs (P-LMXBs) in the model are also shown.}
\label{table:App_model4}      
\centering                        
%%\resizebox{\columnwidth}{!}{ 
\begin{tabular}{c | c c }        % centered columns (4 columns)
\hline\hline                 % inserts double horizontal lines
\noalign{\smallskip}
XRB & \multicolumn{2}{c}{Number of systems}\\    % table heading 
\noalign{\smallskip}
type & \textit{Observed}  & Model 4 \\
\noalign{\smallskip}
\hline                        % inserts single horizontal line
\noalign{\smallskip}
   LMXB & \textit{30}  & 1187 (1.3)  \\ 
  \footnotesize
  P-LMXB & \textit{?}  & \footnotesize 937  \\
  \footnotesize
   oT-LMXB & \textit{?}  & \footnotesize 250 \\
   
   IMXB & \textit{64}  & 94 \, (4.6) \\
   HMXB & \textit{120} & 285 \,  (2.2) \\   
\hline    
\noalign{\smallskip}
Total & \textit{214} & 1566 (8.1) \\
 \smallskip
\end{tabular}
\end{table}

\section{The typical evolution of XRBs which are the progenitors of merging DCOs.}\label{App:evolution}

\begin{figure}
   \centering
   \resizebox{\hsize}{!}
   {\includegraphics[width=\textwidth,angle=270]{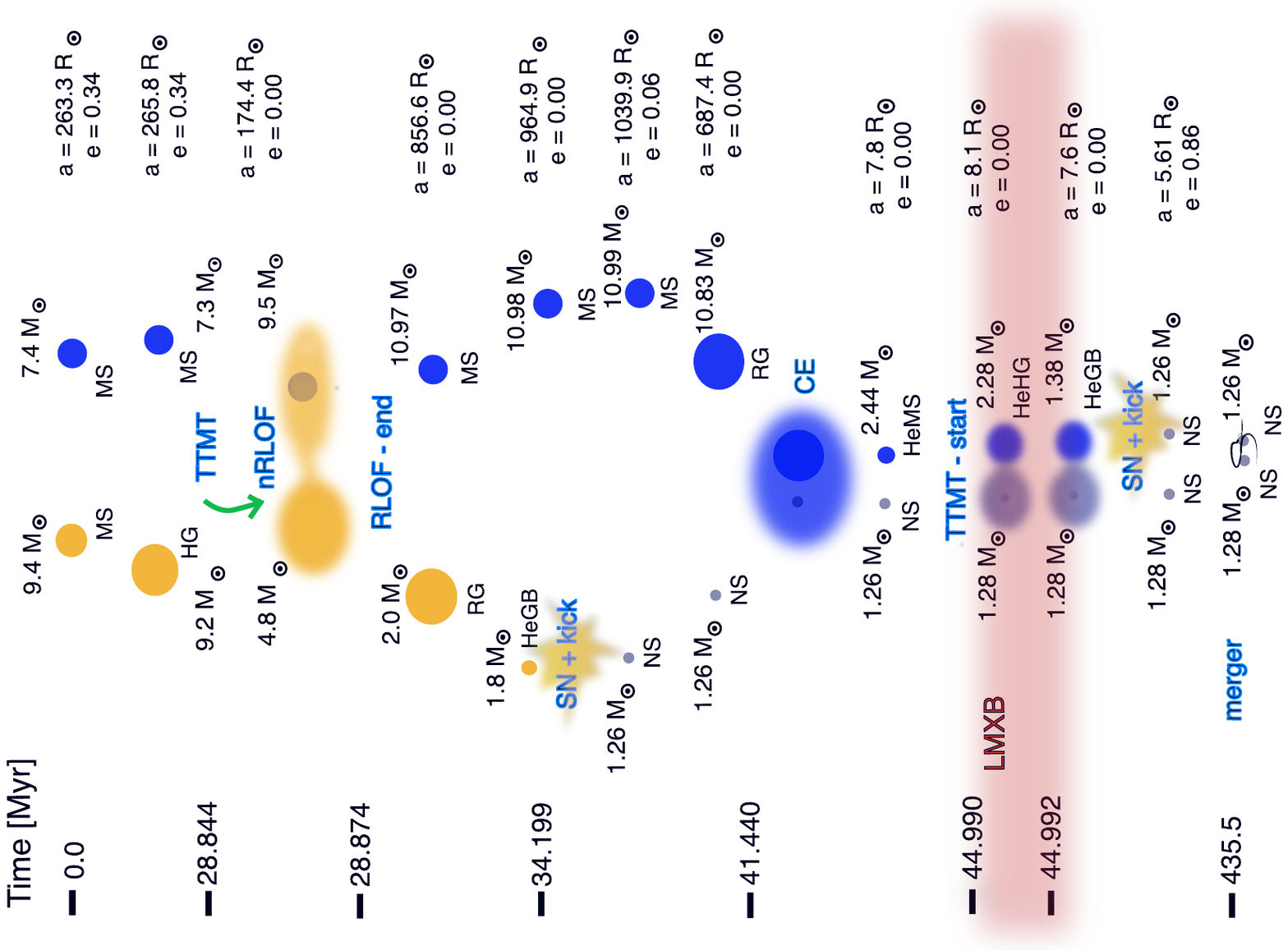}}
 \caption{The typical evolution of the binary that becomes LMXB and ends up as merging NS-NS in Model 1. The time-scale (in Myr) is shown on the left side of the panel, the masses (in $M_{\odot}$) are marked next to the stars they refer to and the orbital separation ($a$ in $R_{\odot}$) and eccentricity ($e$) of the system at the given stage of the evolution are shown on the right side of the panel. The acronyms on the figure stand for:\\
MS – main sequence star\\
HG – Hertzsprung gap star \\
RG – first giant branch star \\
CHeB – core helium burning star \\
HeMS – main sequence naked helium star \\ 
HeHG – Hertzsprung gap naked helium star \\ 
NS – neutron star \\ 
BH – black hole \\
CE - common envelope \\
SN - supernova explosion \\
nRLOF - nuclear Roche lobe overflow \\
TTMT - thermal time-scale mass transfer
}
  \label{Fig:LMXB}
\end{figure}

\begin{figure}
   \centering
   \resizebox{\hsize}{!}
   {\includegraphics[width=\textwidth,angle=270]{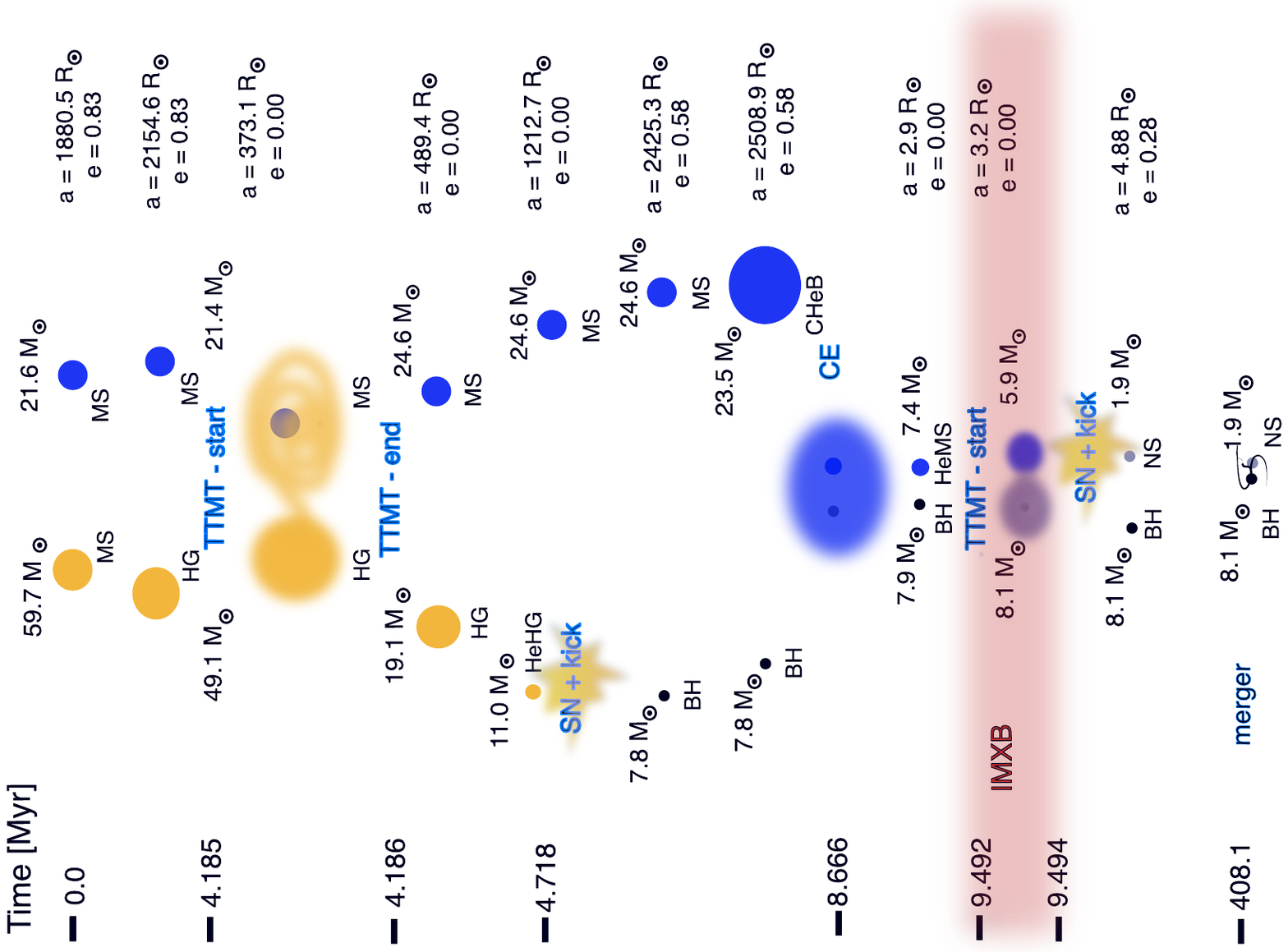}}
 \caption{The typical evolution of the binary that becomes IMXB and ends up as merging NS-BH in Model 1. The time-scale (in Myr) is shown on the left side of the panel, the masses (in $M_{\odot}$) are marked next to the stars they refer to and the orbital separation ($a$ in $R_{\odot}$) and eccentricity ($e$) of the system at the given stage of the evolution are shown on the right side of the panel. The acronyms on the figure are the same as on Fig.~\ref{Fig:LMXB}}.\\
  \label{Fig:IMXB}
\end{figure}

\begin{figure}
   \centering
   \resizebox{\hsize}{!}
   {\includegraphics[width=\textwidth,angle=270]{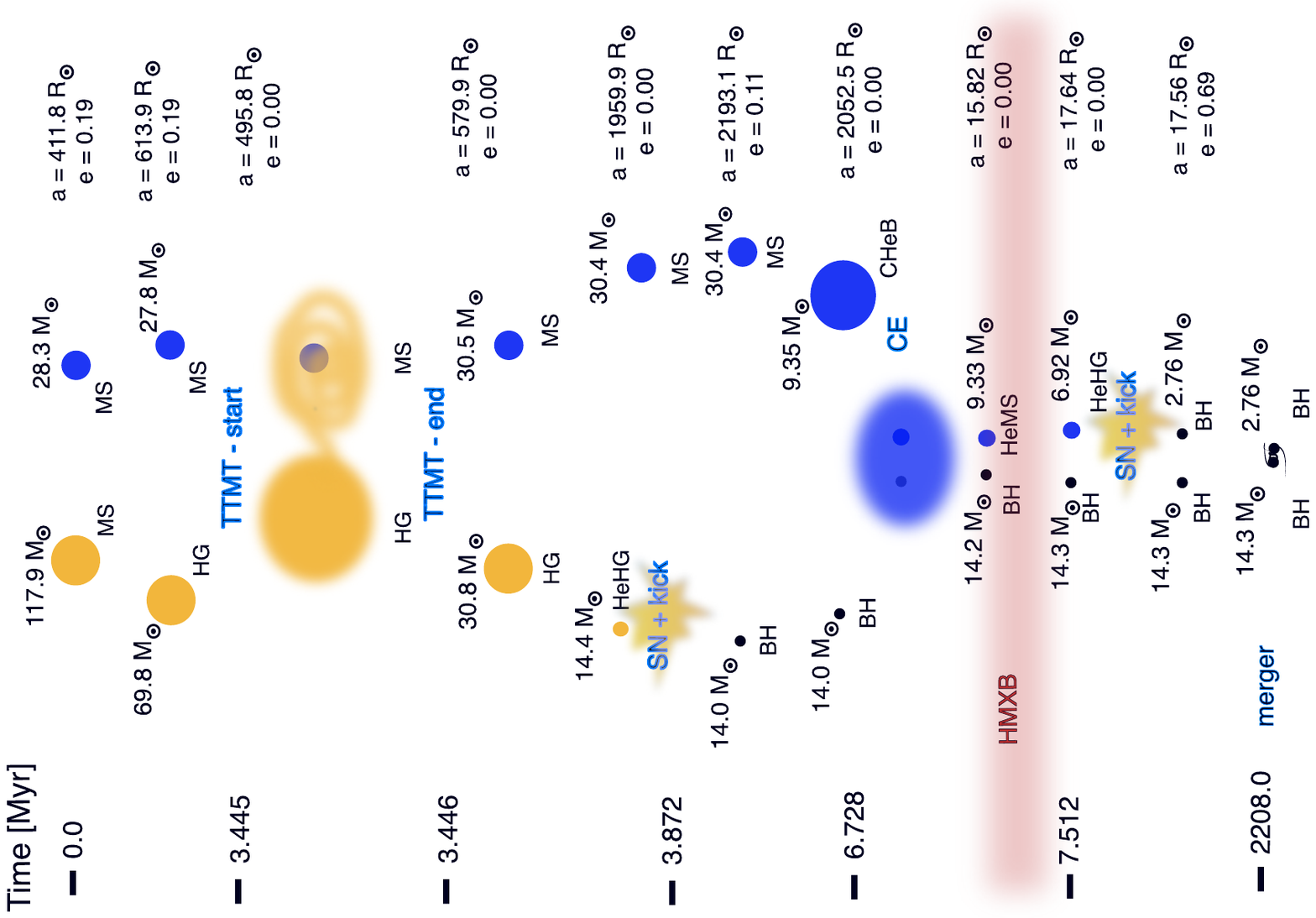}}
 \caption{The typical evolution of the binary that becomes HMXB and ends up as merging BH-BH in Model 1. The time-scale (in Myr) is shown on the left side of the panel, the masses (in $M_{\odot}$) are marked next to the stars they refer to and the orbital separation ($a$ in $R_{\odot}$) and eccentricity ($e$) of the system at the given stage of the evolution are shown on the right side of the panel. The acronyms on the figure are the same as on Fig.~\ref{Fig:LMXB}}.\\
  \label{Fig:HMXB}
\end{figure}

We present the summary of the typical evolution paths of merging DCOs from XRBs in our models in Table~\ref{table:EvolPaths_Model12} and Table~\ref{table:EvolPaths_Model3} for Model 1/2 and Model 3 respectively. We also show the schematic pictures of the typical evolutionary scenarios leading to merging BH-BH from HMXB (Fig.~\ref{Fig:HMXB}, BH-NS from IMXB (Fig.~\ref{Fig:IMXB} in Model 1 and 2 and NS-NS from LMXB (Fig.~\ref{Fig:LMXB} in Model 1. More detailed description can be found in main text in Sec.~\ref{subsec:Discussion_mergingDCOs}.

\begin{table*}
\caption{
The dominant formation scenarios of the merging DCOs in \textbf{Model 1} and \textbf{Model 2}. Note that the formation of NS-NS from LMXBs applies to Model 1 only.}
\label{table:EvolPaths_Model12}      
\centering                        
%%\resizebox{\columnwidth}{!}{ 
\begin{tabular}{c c c }        % centered columns (4 columns)
\hline\hline                 % inserts double horizontal lines
\noalign{\smallskip}
DCO type & \textit{Forming scenario}  & XRB progenitor type \\
\noalign{\smallskip}
\hline                        % inserts single horizontal line
\noalign{\smallskip}
BH-BH & MT1(2-1) BH1 CE2(14-4;14-7) BH2  & HMXB \\    
BH-NS & MT1(2/9-1) BH1 CE2(14-4;14-7) MT2(14-8) NS2 & IMXB \\
NS-NS & MT1(2/3/9-1) NS1 CE2(13-5;13-8) MT2(13-8/9) NS2 & HMXB \\   
NS-NS & MT(2/8/9-1) NS1 CE2(13-4;13-7) MT2(13-8/9) NS2 & LMXB \\
\hline    
 \smallskip
\end{tabular}
\tablefoot{\\
MT1-stable RLOF, donor is initially more massive star;\\
MT2-stable RLOF, donor is initially less massive star;\\
BH1-formation of black hole by initially more massive star;\\
BH2-formation of black hole by initially less massive star;\\
NS1-formation of neutron star by initially more massive star;\\
NS2-formation of neutron star by initially less massive star;\\
\smallskip
CE2-common envelope initiated by initially less massive star;\\
Number following either ”CE” or ”MT” marks the donor star, ”1” stands for initially more massive star, ”2” stands initially less massive star.\\
The first value in the brackets is the type of an initially more massive component and the second value is the type of an initially less massive component. The values after slashes, e.g. 2/9, refer to the different possible evolutionary types of binary components during
given phase. In the case of CE phase the colon divides the components types at the onset (before colon) from the components types after the envelope ejection (after colon).\\
The numeric types are consistent with Hurley et al. (2002):\\
1 – main sequence star with $M>0.7\,M_{\odot}$\\
2 – Hertzsprung gap star \\
3 – first giant branch star \\
4 – core helium burning star \\
5 – early asymptotic giant branch star \\ 
7 – main sequence naked helium star \\ 
8 – Hertzsprung gap naked helium star \\ 
9 – giant branch naked helium star \\ 
13 – neutron star \\ 
14 – black hole \\ 
}
\end{table*}

\begin{table*}
\caption{
The dominant formation scenarios of the merging DCOs in \textbf{Model 3}.}
\label{table:EvolPaths_Model3}      
\centering                        
%%\resizebox{\columnwidth}{!}{ 
\begin{tabular}{c c c }        % centered columns (4 columns)
\hline\hline                 % inserts double horizontal lines
\noalign{\smallskip}
DCO type & \textit{Forming scenario}  & XRB progenitor type  \\
\noalign{\smallskip}
\hline                        % inserts single horizontal line
\noalign{\smallskip}
BH-BH & MT1(2-1) BH1 MT2(14-2) BH2  & HMXB \\    
BH-NS & MT1(2-1) BH1 CE2(14-4;14-7) MT2(14-8) NS2 & IMXB \\
NS-NS & MT1(5/6-1) NS1 CE2(13-5;13-8) MT2(13-8/9) NS2 & HMXB \\
\hline    
 \smallskip
\end{tabular}
\end{table*}

\end{appendix}

\end{document}